\documentclass[aps,prb,twocolumn,groupedaddress,showpacs,floatfix,amsmath]{revtex4}
\def\bc{\begin{center}}
\def\ec{\end{center}}
\def\be{\begin{equation}}
\def\ee{\end{equation}}
\renewcommand{\vec}[1]{\mbox{\boldmath$#1$}}
\usepackage{epsfig}
\usepackage{dcolumn}

\begin{document}

\title{Semiconductor quantum dots in high magnetic fields: The composite-fermion
view}
\author{Gun Sang Jeon,$^{1,2}$ Chia-Chen Chang,$^1$ and Jainendra K. Jain$^1$}
\affiliation{$^1$Department of Physics, 104 Davey Laboratory, The Pennsylvania State University,
University Park, Pennsylvania 16802, USA}
\affiliation{$^2$Department of Physics and Astronomy, Seoul National University,
Seoul 151-747, Korea}

\date{\today}

\begin{abstract}
We review and extend the composite fermion theory for semiconductor quantum dots in high magnetic fields. The mean-field model of composite fermions is 
unsatisfactory for the qualitative physics at high angular momenta.
Extensive numerical calculations demonstrate that the microscopic CF theory, 
which incorporates interactions between composite fermions, provides an excellent qualitative and quantitative account 
of the quantum dot ground state down to the largest angular momenta studied, 
and allows systematic improvements by inclusion of mixing between composite
fermion Landau levels (called $\Lambda$ levels).  

\end{abstract}
\maketitle

\section{Introduction}

The system of interacting electrons confined to a two dimensional 
quantum dot and exposed to a strong magnetic field has been a subject 
of intense theoretical study for over two 
decades.~\cite{review,Yoshioka,Girvin,Dev,Beenakker93,Yang,Tapash,Xie,Hawrylak,
Kawamura,Kamilla,Seki,Manninen,Cappelli,Landman,Ruan,Maksym,Harju,RQLC,
Rei2, Toreblad, Cnell,Rei5,Peeters2,Muller}
Such quantum dots have been realized and studied in 
the laboratory.~\cite{Expt1,Expt2,Expt3,Expt4}
Exact diagonalization studies show that the ground states are strongly 
correlated, and the aim of theory is to achieve a satisfactory understanding of the 
correlations.  It is also of interest to understand how this 
ties into our understanding of the FQHE,~\cite{Tsui} obtained in the 
thermodynamic limit without confinement.

The CF theory has been applied to parabolic quantum dots subjected to a 
strong magnetic field.  The plot of ground state energy as 
a function of the angular momentum ($L$) has a rich structure.
In particular, downward cusps appear at certain values of $L$, 
which are consequently especially favorable. 
Early studies~\cite{Dev,Beenakker93,Kawamura,Kamilla} demonstrated 
the CF theory to be promising.  Specifically, 
a ``mean-field model," in which the composite fermions
are taken as noninteracting particles at an effective angular momentum
$L^*$, with their mass or the cyclotron energy
treated as a phenomenological parameter, predicts cusps
in the energy at certain magic angular momenta; these predictions are 
in agreement with the actual cusp positions in exact diagonalization studies 
at small angular momenta $L$, but discrepancies appear at large 
$L$.~\cite{Seki,Landman}  Further work~\cite{Jeon,Jeon2} showed 
that these discrepancies are special to the mean-field model of the CF theory.
A perfect agreement between the actual and the predicted cusp positions 
was obtained when the CF energies were calculated from microscopic wave functions.
One of the surprising aspects was the success of the CF theory even  
at the largest angular momenta studied, which 
appears, at first, to be at odds with the 
classical crystal-like correlations found in exact diagonalization 
studies.~\cite{Seki,Maksym}  While both composite fermions and the 
crystal are generated by the repulsive interaction between electrons, 
the implicit assumption had been that one excluded the other.  
The work in Ref.~\onlinecite{Jeon2} showed that 
no logical inconsistency exists between the simultaneous 
formations of composite fermions and crystal-like structures at 
low fillings, and, furthermore, the formation of composite fermions itself induces  
crystal structure at low fillings.  This crystal has been shown to be very well described 
as a crystal of composite fermions.~\cite{CFcrystal,onefifth}

The aim of this paper is to review and extend the CF theory of  
quantum dots in high magnetic fields, and also provide many
 details left out in earlier papers.  
Section II briefly outlines the basics of the CF theory for quantum dot states, giving 
explicit wave functions for some simple cases.
Section III describes the numerical methods (exact diagonalization, Lanczos, and 
CF diagonalization).  The mean-field CF model is discussed in Sec.~IV, and the 
``zeroth-order" CF diagonalization in Sec.~V.  Section VI illustrates how the 
results are improved by going to higher orders in the CF theory.  
The paper is concluded in Sec.~VII.

\section{The composite-fermion basics for quantum dots}

Following the standard practice, we assume below parabolic confinement. 
This should be a good approximation for most quantum dots at low energies, 
and simplifies calculations because of the availability of exact solutions for 
single particle eigenstates. The Hamiltonian of interest is  
\be
H{=}\sum_j\frac{1}{2m_b}\left(\vec{p}_j+\frac{e}{c}\vec{A}_j\right)^2
+\sum_j\frac{m_b}{2} \omega_0^2 r_j^2 + 
\sum_{j<k} \frac{e^2}{\epsilon r_{jk}},
\ee
where $m_b$ is the band mass of the electron, $\omega_0$ 
is a measure of the strength of the confinement, $\epsilon$ is 
the dielectric constant of the host semiconductor, and $r_{jk}\equiv
|\vec{r}_j-\vec{r}_k|$. 
We will specialize to the case of very large magnetic fields,
when $\omega_c \equiv eB/m_b c \gg \omega_0$.  Only the lowest Landau level
(LL) is relevant in this limit.  In that limit, at each angular momentum the 
eigenenergy neatly separates into a confinement part and an 
interaction part:
\be
E(L)=E_c(L) + V(L),
\ee
where 
$E_{c}(L)=(\hbar/2)[\Omega-\omega_{c}] L$,
relative to the lowest LL, with 
$\Omega^2\equiv \omega_{c}^2+4\omega_{0}^2$,
and $V(L)$ is the interaction energy of electrons without 
confinement, but with the magnetic length replaced by 
an effective magnetic length given by $\ell\equiv
\sqrt{\hbar/m_b\Omega}$. 
In the following, we will consider only $V$ as a function of the 
angular momentum $L$; it must be remembered, however, 
that the confinement part must be added to determine the 
global ground state.

\begin{figure}
\centerline{\epsfig{file=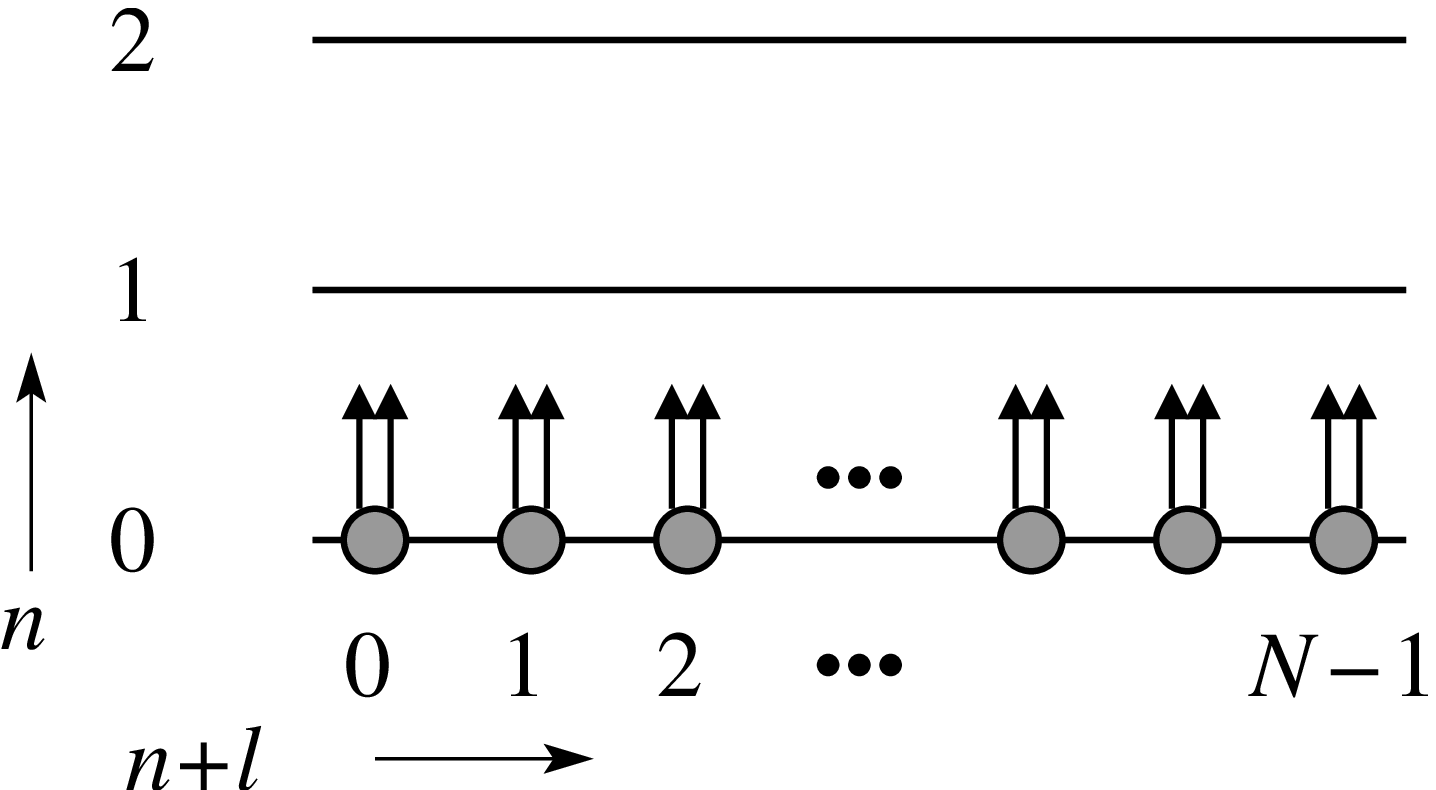,width=0.4\textwidth,angle=0}}
\centerline{(a)}\vspace*{5mm}
\centerline{\epsfig{file=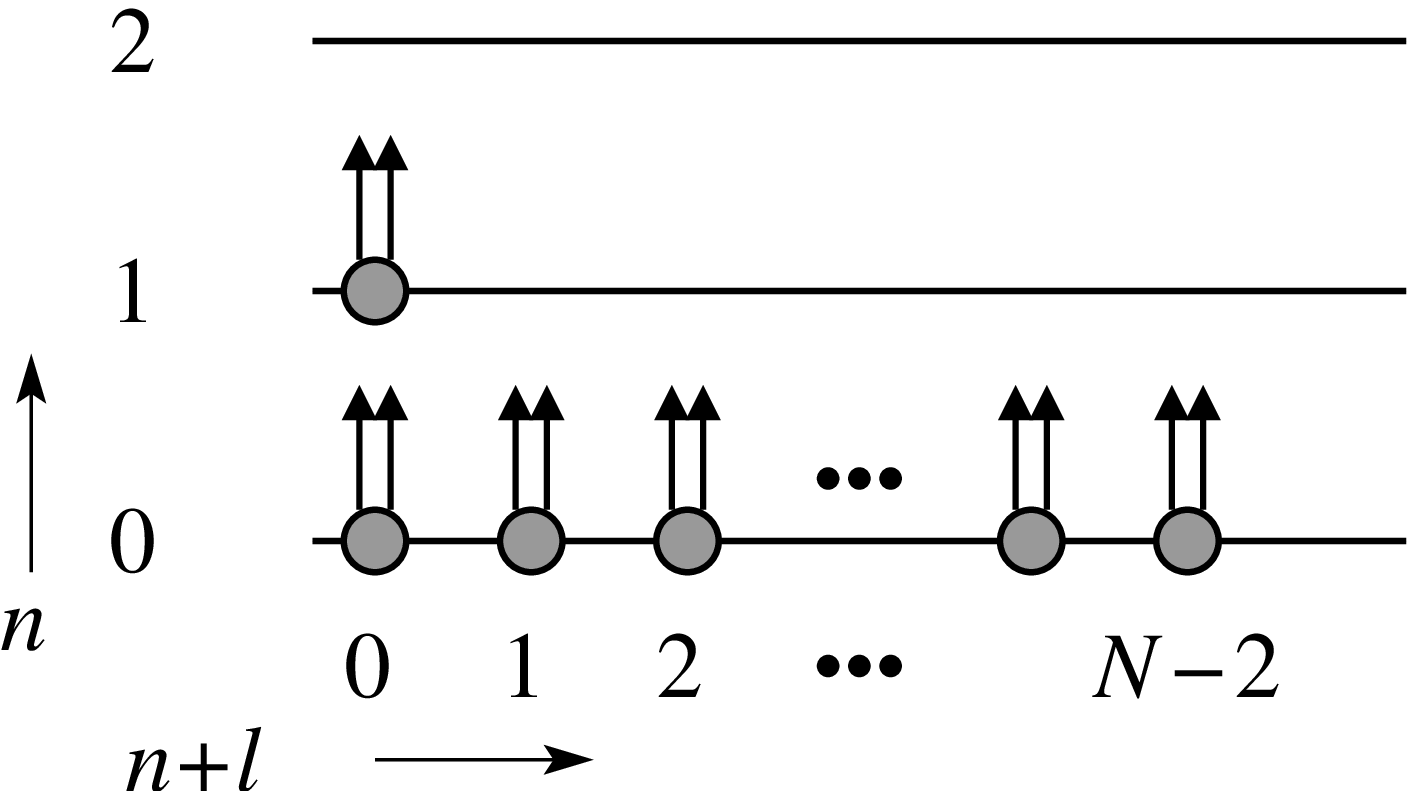,width=0.4\textwidth,angle=0}}
\centerline{(b)}\vspace*{5mm}
\centerline{\epsfig{file=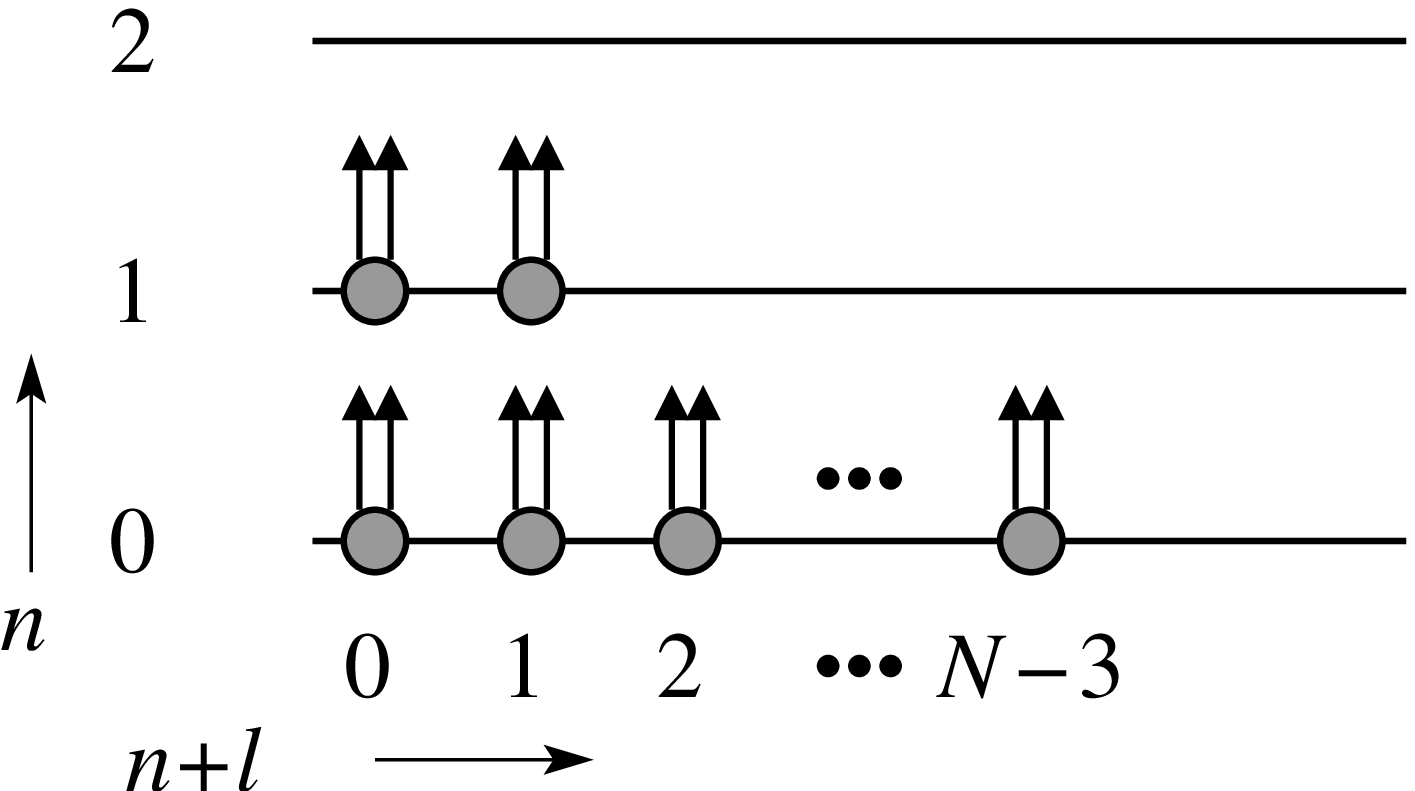,width=0.4\textwidth,angle=0}}
\centerline{(c)}\vspace*{5mm}
\caption{ \label{fig:CFlevel}
Schematic depiction of (a) $[N,0]$, 
(b) $[N-1,1]$, and (c) $[N-2,2]$ states
in terms of composite fermions carrying two vortices ($2p=2$).
The composite fermions are shown as electrons with two arrows, 
where the arrows represent the vortices bound to composite
fermions.  The labels 
$n$ and $l$ denote the $\Lambda$ level (CF Landau level) 
index and the angular
momentum of the composite fermion, respectively.
}
\end{figure}

In the CF theory,~\cite{Dev,Kawamura,Jain} the interacting state of electrons 
in the lowest LL at angular momentum $L$ is 
mapped into the noninteracting electron state at $L^*$,
where
\begin{equation}
L=L^*+pN(N-1)\;,
\end{equation}
$N$ is the number of electrons, and $p$ is an integer.  
The wave functions
\begin{equation}
\Psi^L_\alpha={\cal P}_{\rm LLL} \prod_{j<k}(z_j-z_k)^{2p} \Phi^{L^*}_\alpha 
\end{equation}
give ansatz wave functions for interacting electrons at $L$ in terms of
the known wave functions of noninteracting electrons at $L^*$.
Here $\Phi^{L^*}_\alpha$ are the wave functions for noninteracting 
electrons at $L^*$ (which in general occupy 
several Landau levels), $\alpha=1, 2, \cdots, D^*$ labels the different states,
$z_j=x_j-iy_j$ denotes the position of the $j$th electron, 
${\cal P}_{\rm LLL}$ indicates projection into the lowest LL, 
$\Psi_\alpha^L$ are basis functions for interacting electrons 
at $L$, and $D^*$ is the dimension of the 
CF basis.  We will restrict $\Phi_\alpha^{L^*}$ to states with the lowest
kinetic energy at $L^*$, and choose $p$ so as to have the smallest 
dimension for the basis.  The composite fermions carrying $2p$ vortices
are labeled $^{2p}$CF's.
At certain values of $L$, the above prescription produces only one 
state ($D^*=1$), which is the CF theory's answer for the 
ground state.  In the notation of 
Ref.~\onlinecite{Kawamura}, this is a compact state, denoted by $[N_0,N_1,\cdots]$,
with $N_j$ composite fermions compactly occupying the innermost angular momentum
orbitals of the $j$th CF level.  At other values of $L$, 
when there are many CF basis states ($D^*>1$), we diagonalize the 
Coulomb Hamiltonian in the CF basis to obtain the ground state,
using methods described earlier.~\cite{Kamilla,Mandal}
For any $N$, there are many values of $L$ where the CF theory 
gives a unique answer, but in general, $D^*$ increases with $N$.

\subsection{Examples of CF bases}

Construction of the CF basis is in principle straightforward.
We consider some explicit examples.

In many cases, the CF basis consists of a unique wave function, which
simplifies the analysis tremendously.  
The simplest wave function is the one for one filled
$\Lambda$ level (CF Landau level), shown schematically in Fig.~\ref{fig:CFlevel}(a):
\begin{eqnarray} \nonumber
\Psi &=&
A\left[\prod_{i=1}^{N}  z_i^{i-1} \right] \Phi_1^{2p}
\\
&=&\prod_{j<k}(z_j-z_k)^{2p+1} e^{-\sum_{l=1}^N |z_l|^2/4} ,
\label{wf1}
\end{eqnarray}
where $A$ denotes an antisymmetric Slater determinant, i.e.
\begin{eqnarray} 
\nonumber
A\left[\prod_{i=1}^{N}  z_i^{i-1} \right] &=&
\left|
\begin{matrix}
1 & 1 & \ldots & 1\\
z_1 & z_2 & \ldots & z_N\\
\vdots & \vdots&\ddots & \vdots\\
z_1^{N-1} & z_2^{N-1} & \ldots & z_N^{N-1}
\end{matrix}
\right|
\\
		&=& \prod_{j<k} (z_j - z_k)
\end{eqnarray}
and
\begin{equation}
\Phi_1^{2p} \equiv
\prod_{j<k}(z_j-z_k)^{2p} e^{-\sum_{l=1}^N |z_l|^2/4} .
\end{equation}
It corresponds to $L=(2p+1)N(N-1)/2$.
It is identical to the Laughlin wave function.

The state $[N-1,1]$
at $L=(2p+1)N(N-1)/2-N$, shown in Fig.~\ref{fig:CFlevel}(b), represents  
the wave function
\begin{equation} \label{wf2}
\Psi = {\cal P}_{\rm LLL} A\left[ z_{1}^*\cdot
\prod_{i=2}^{N}  z_i^{i-2} \right]
\Phi_1^{2p} .
\end{equation}
This state has been interpreted as a charged quasiparticle excitation
of the $\nu=1/(2p+1)$ FQHE state.~\cite{QP}
The $[N-2,2]$ state (Fig.~\ref{fig:CFlevel}(c)) occurs at
$L=(2p+1)N(N-1)/2-2(N-1)$, and has the wave function
\begin{equation} \label{wf3}
\Psi = {\cal P}_{\rm LLL} A\left[ z_{1}^*\cdot z_{2} z_{2}^*\cdot
\prod_{i=3}^{N}  z_i^{i-3} \right]
\Phi_1^{2p}.
\end{equation}

\begin{figure}
\parbox{0.2\textwidth}{
\epsfig{file=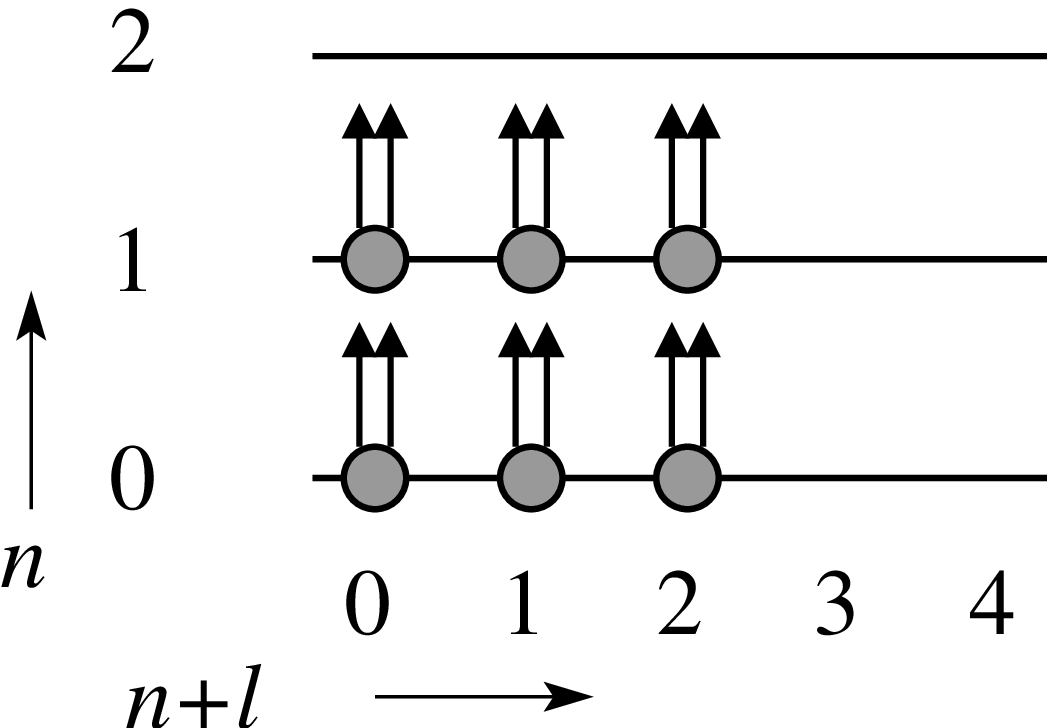,width=0.2\textwidth}
}
\hspace*{0.05\textwidth}
\parbox{0.2\textwidth}{
\epsfig{file=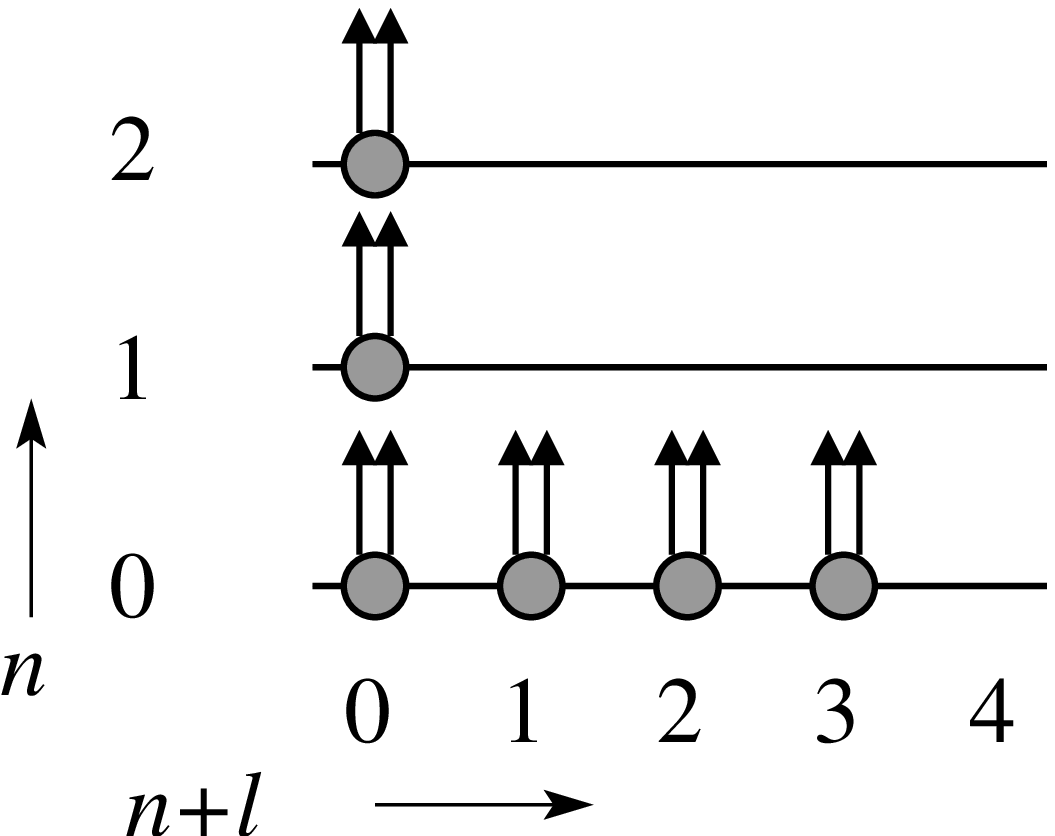,width=0.2\textwidth}
}
\centerline{(a)}\vspace*{5mm}
\parbox{0.2\textwidth}{
\epsfig{file=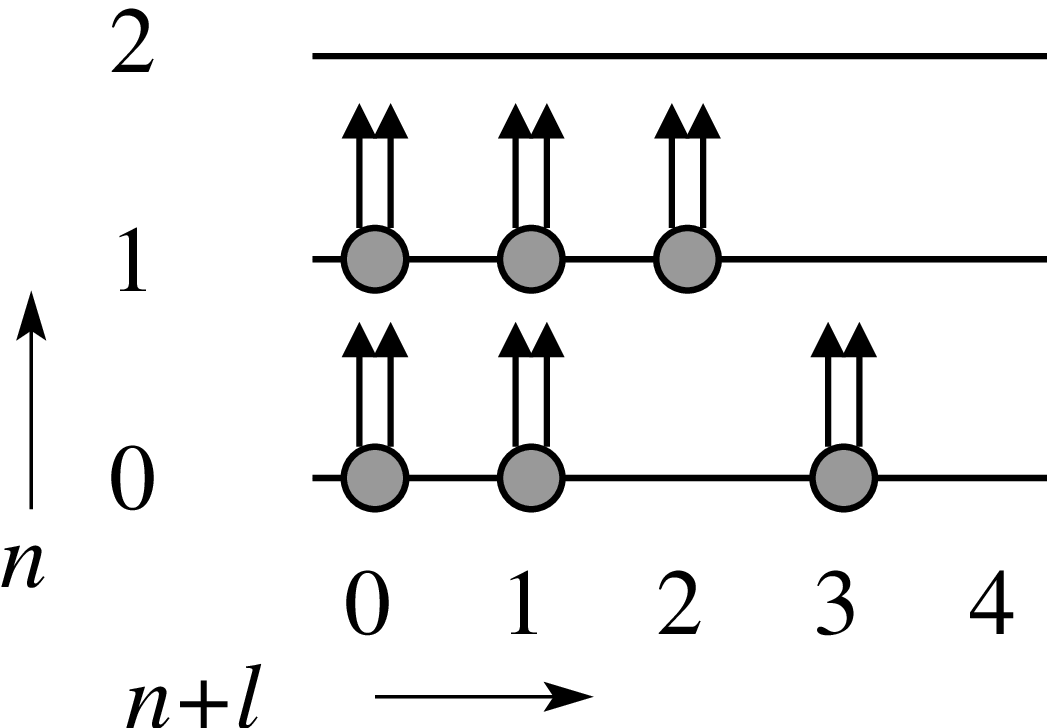,width=0.2\textwidth}
}
\hspace*{0.05\textwidth}
\parbox{0.2\textwidth}{
\epsfig{file=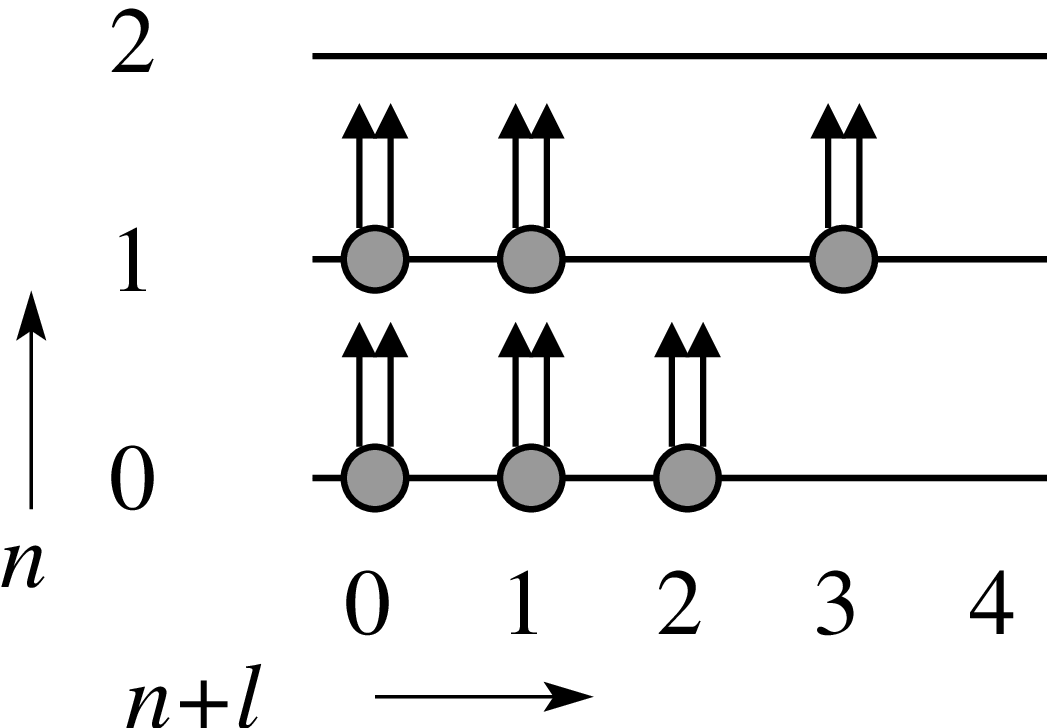,width=0.2\textwidth}
}\\
\parbox{0.2\textwidth}{
\epsfig{file=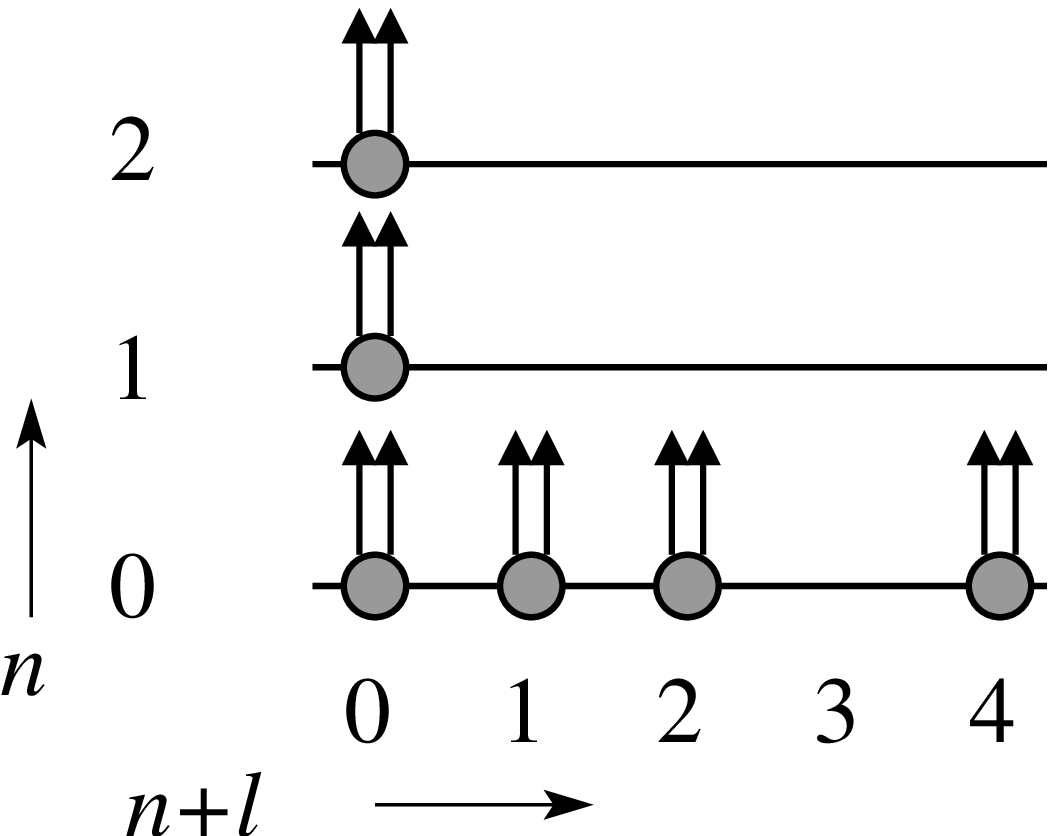,width=0.2\textwidth}
}
\hspace*{0.05\textwidth}
\parbox{0.2\textwidth}{
\epsfig{file=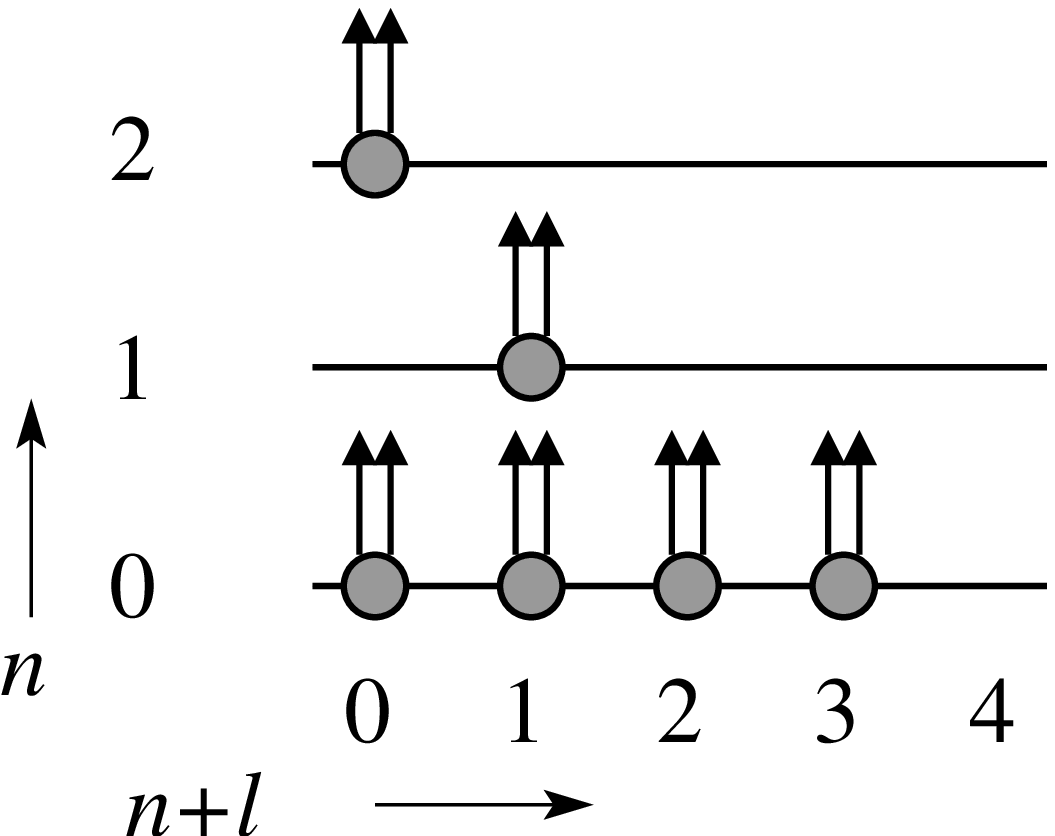,width=0.2\textwidth}
}\\
\parbox{0.2\textwidth}{
\epsfig{file=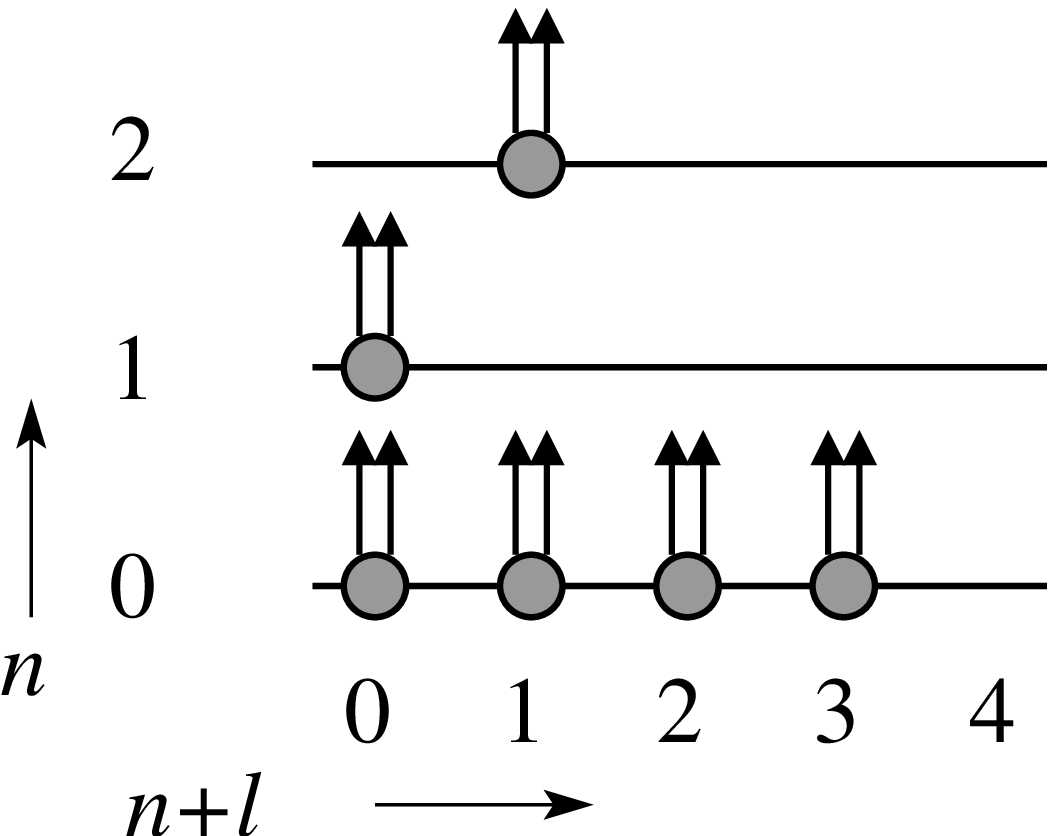,width=0.2\textwidth}
}
\hspace*{0.25\textwidth}$\phantom{a}$
\\
\centerline{(b)}
\caption{\label{fig:CFlevel2}
Schematic depiction of basis states at (a) $L=33$; (b) $L=34$
for $N=6$ particles.
}
\end{figure}

For many angular momenta, the basis contains more than one state.
For six composite fermions at $L=33$ we have two compact states [3,3] and [4,1,1], as depicted 
schematically in Fig.~\ref{fig:CFlevel2}(a).
The basis functions are given by
\begin{subequations}
\begin{eqnarray}
\Psi_1^{L=33} &=& {\cal P}_{\rm LLL} A
\left[ \prod_{i=1}^3 \left(z_{i}^* z_{i}^{i-1} \right)
\cdot
\prod_{i=4}^{6}  z_i^{i-4} \right]
\Phi_1^{2},
\\
\Psi_2^{L=33} &=& {\cal P}_{\rm LLL} A
\left[ (z_1^*)^2 \cdot z_2^*
\cdot
\prod_{i=3}^{6}  z_i^{i-3} \right]
\Phi_1^{2}.
\end{eqnarray}
\end{subequations}
These have the same CF kinetic energy, leading to $D^*=2$ at $L=33$.
The five basis states at $L=34$ are shown in Fig.~\ref{fig:CFlevel2}(b), derived 
from either [4,1,1] or [3,3] by increasing the angular momentum by one unit.
The corresponding basis functions are written as
\begin{widetext}
\begin{subequations}
\begin{eqnarray}
\Psi_1^{L=34} &=& {\cal P}_{\rm LLL} A \left[
\eta_1^{-1}(z_1) \cdot \eta_1^0(z_2) \cdot \eta_1^1(z_3) 
\cdot \eta_0^0(z_4)\cdot \eta_0^1(z_5)\cdot \eta_0^3(z_6)
\right]
\Phi_1^{2} ,
\\
\Psi_2^{L=34} &=& {\cal P}_{\rm LLL} A \left[
\eta_1^{-1}(z_1) \cdot \eta_1^0(z_2) \cdot \eta_1^2(z_3) 
\cdot \eta_0^0(z_4)\cdot \eta_0^1(z_5)\cdot \eta_0^2(z_6)
\right]
\Phi_1^{2} ,
\\
\Psi_3^{L=34} &=& {\cal P}_{\rm LLL} A \left[
\eta_2^{-2}(z_1) \cdot \eta_1^{-1}(z_2) \cdot \eta_0^0(z_3) 
\cdot \eta_0^1(z_4)\cdot \eta_0^2(z_5)\cdot \eta_0^4(z_6)
\right]
\Phi_1^{2} ,
\\
\Psi_4^{L=34} &=& {\cal P}_{\rm LLL} A \left[
\eta_2^{-2}(z_1) \cdot \eta_1^0(z_2) \cdot \eta_0^0(z_3) 
\cdot \eta_0^1(z_4)\cdot \eta_0^2(z_5)\cdot \eta_0^3(z_6)
\right]
\Phi_1^{2} ,
\\
\Psi_5^{L=34} &=& {\cal P}_{\rm LLL} A \left[
\eta_2^{-1}(z_1) \cdot \eta_1^{-1}(z_2) \cdot \eta_0^0(z_3) 
\cdot \eta_0^1(z_4)\cdot \eta_0^2(z_5)\cdot \eta_0^3(z_6)
\right]
\Phi_1^{2} .
\end{eqnarray}
\end{subequations}
\end{widetext}
Here
\begin{equation}
\eta_n^l(z) \equiv z^l L_n^l (z z^*/2) 
\end{equation}
with $L_n^l(x)$ being an associated Laguerre polynomial.

\begingroup
\squeezetable
\begin{table*}
\begin{ruledtabular}
\begin{tabular}{r r r l l | r r r l l| r r r l l| r r r l l}
 $L$ &  $D$ & $D^*$ &  $V_{\rm ex}$ &  $V_{\rm CF}$ &  $L$ &  $D$ & $D^*$ &  $V_{\rm
 ex}$ &  $V_{\rm CF}$ &   $L$ &  $D$ & $D^*$ &  $V_{\rm ex}$ &  $V_{\rm CF}$ &  $L$ &
 $D$ & $D^*$ &  $V_{\rm ex}$ &  $V_{\rm CF}$ \\ 
\hline 
 9 &  3 &  1 &  1.93481 &  1.93462(17) &  39 &  378 &  2 &  0.88879 &  0.88986(5) &  69 &  2178 &  2 &  0.66433 &  0.66487(3) &  99 &  6528 &  2 &  0.54944 &  0.54965(1) \\ 
 10 &  5 &  1 &  1.78509 &  1.78496(21) &  40 &  411 &  4 &  0.88032 &  0.88147(7) &  70 &  2280 &  1 &  0.65131 &  0.65168(2) &  100 &  6736 &  4 &  0.54748 &  0.54790(3) \\ 
 11 &  6 &  2 &  1.78509 &  1.78487(21) &  41 &  441 &  6 &  0.87275 &  0.87410(5) &  71 &  2376 &  2 &  0.65131 &  0.65167(4) &  101 &  6936 &  6 &  0.54581 &  0.54625(4) \\ 
 12 &  9 &  1 &  1.68518 &  1.68616(7) &  42 &  478 &  1 &  0.84446 &  0.84822(7) &  72 &  2484 &  1 &  0.64803 &  0.65220(4) &  102 &  7153 &  1 &  0.53846 &  0.53874(2) \\ 
 13 &  11 &  2 &  1.64157 &  1.64407(27) &  43 &  511 &  1 &  0.84446 &  0.84813(9) &  73 &  2586 &  2 &  0.64520 &  0.64678(7) &  103 &  7361 &  1 &  0.53846 &  0.53879(3) \\ 
 14 &  15 &  1 &  1.50066 &  1.50174(20) &  44 &  551 &  2 &  0.83722 &  0.84177(10) &  74 &  2700 &  1 &  0.63324 &  0.63361(3) &  104 &  7586 &  2 &  0.53662 &  0.53713(3) \\ 
 15 &  18 &  2 &  1.50066 &  1.50157(7) &  45 &  588 &  2 &  0.83079 &  0.83160(12) &  75 &  2808 &  2 &  0.63324 &  0.63362(2) &  105 &  7803 &  2 &  0.53505 &  0.53527(2) \\ 
 16 &  23 &  4 &  1.46397 &  1.46424(28) &  46 &  632 &  1 &  0.80616 &  0.80683(7) &  76 &  2928 &  4 &  0.63023 &  0.63088(4) &  106 &  8037 &  1 &  0.52812 &  0.52825(2) \\ 
 17 &  27 &  6 &  1.42958 &  1.42999(43) &  47 &  672 &  2 &  0.80616 &  0.80680(8) &  77 &  3042 &  6 &  0.62764 &  0.62843(1) &  107 &  8262 &  2 &  0.52812 &  0.52825(4) \\ 
 18 &  34 &  1 &  1.30573 &  1.31078(10) &  48 &  720 &  1 &  0.79987 &  0.80871(4) &  78 &  3169 &  1 &  0.61660 &  0.61727(2) &  108 &  8505 &  1 &  0.52638 &  0.52761(3) \\ 
 19 &  39 &  1 &  1.30573 &  1.31054(25) &  49 &  764 &  2 &  0.79434 &  0.79747(11) &  79 &  3289 &  1 &  0.61660 &  0.61732(5) &  109 &  8739 &  2 &  0.52490 &  0.52543(2) \\ 
 20 &  47 &  2 &  1.27825 &  1.28375(7) &  50 &  816 &  1 &  0.77263 &  0.77345(10) &  80 &  3422 &  2 &  0.61382 &  0.61499(3) &  110 &  8991 &  1 &  0.51834 &  0.51850(2) \\ 
 21 &  54 &  2 &  1.24416 &  1.24456(13) &  51 &  864 &  2 &  0.77263 &  0.77337(5) &  81 &  3549 &  2 &  0.61143 &  0.61186(7) &  111 &  9234 &  2 &  0.51834 &  0.51850(2) \\ 
 22 &  64 &  1 &  1.17779 &  1.17793(7) &  52 &  920 &  4 &  0.76711 &  0.76815(4) &  82 &  3689 &  1 &  0.60120 &  0.60147(2) &  112 &  9495 &  4 &  0.51670 &  0.51700(4) \\ 
 23 &  72 &  2 &  1.17779 &  1.17792(9) &  53 &  972 &  6 &  0.76229 &  0.76350(7) &  83 &  3822 &  2 &  0.60120 &  0.60146(3) &  113 &  9747 &  6 &  0.51530 &  0.51570(2) \\ 
 24 &  84 &  1 &  1.15660 &  1.16365(10) &  54 &  1033 &  1 &  0.74297 &  0.74506(4) &  84 &  3969 &  1 &  0.59863 &  0.60135(5) &  114 &  10018 &  1 &  0.50910 &  0.50934(4) \\ 
 25 &  94 &  2 &  1.13775 &  1.14169(19) &  55 &  1089 &  1 &  0.74297 &  0.74507(2) &  85 &  4109 &  3 &  0.59642 &  0.59747(5) &  115 &  10279 &  1 &  0.50910 &  0.50934(2) \\ 
 26 &  108 &  1 &  1.08038 &  1.08168(7) &  56 &  1154 &  2 &  0.73807 &  0.74093(4) &  86 &  4263 &  1 &  0.58690 &  0.58720(5) &  116 &  10559 &  2 &  0.50754 &  0.50795(3) \\ 
 27 &  120 &  2 &  1.08038 &  1.08170(19) &  57 &  1215 &  2 &  0.73381 &  0.73440(4) &  87 &  4410 &  2 &  0.58690 &  0.58719(6) &  117 &  10830 &  2 &  0.50622 &  0.50640(1) \\ 
 28 &  136 &  4 &  1.06508 &  1.06589(12) &  58 &  1285 &  1 &  0.71648 &  0.71693(6) &  88 &  4571 &  4 &  0.58451 &  0.58502(4) &  118 &  11120 &  1 &  0.50033 &  0.50043(2) \\ 
 29 &  150 &  6 &  1.05089 &  1.05181(17) &  59 &  1350 &  2 &  0.71648 &  0.71700(10) &  89 &  4725 &  6 &  0.58246 &  0.58306(4) &  119 &  11400 &  2 &  0.50033 &  0.50045(5) \\ 
 30 &  169 &  1 &  1.00340 &  1.00915(10) &  60 &  1425 &  1 &  0.71209 &  0.71838(9) &  90 &  4894 &  1 &  0.57358 &  0.57404(2) &  120 &  11700 &  1 &  0.49885 &  0.49971(2) \\ 
 31 &  185 &  1 &  1.00340 &  1.00922(11) &  61 &  1495 &  2 &  0.70829 &  0.71053(5) &  91 &  5055 &  1 &  0.57358 &  0.57402(3) &  121 &  11990 &  2 &  0.49759 &  0.49801(4) \\ 
 32 &  206 &  2 &  0.99110 &  0.99756(13) &  62 &  1575 &  1 &  0.69263 &  0.69317(2) &  92 &  5231 &  2 &  0.57135 &  0.57208(4) &  122 &  12300 &  1 &  0.49199 &  0.49212(1) \\ 
 33 &  225 &  2 &  0.97949 &  0.98037(14) &  63 &  1650 &  2 &  0.69263 &  0.69316(4) &  93 &  5400 &  2 &  0.56944 &  0.56976(3) &  123 &  12600 &  2 &  0.49199 &  0.49211(2) \\ 
 34 &  249 &  1 &  0.94091 &  0.94152(6) &  64 &  1735 &  4 &  0.68868 &  0.68955(3) &  94 &  5584 &  1 &  0.56112 &  0.56130(1) &  124 &  12920 &  4 &  0.49059 &  0.49084(2) \\ 
 35 &  270 &  2 &  0.94091 &  0.94159(10) &  65 &  1815 &  6 &  0.68526 &  0.68631(6) &  95 &  5760 &  2 &  0.56112 &  0.56132(2) &  125 &  13230 &  6 &  0.48940 &  0.48969(4) \\ 
 36 &  297 &  1 &  0.93079 &  0.94124(14) &  66 &  1906 &  1 &  0.67102 &  0.67223(3) &  96 &  5952 &  1 &  0.55903 &  0.56084(3) &  126 &  13561 &  1 &  0.48406 &  0.48423(2) \\ 
 37 &  321 &  2 &  0.92164 &  0.92557(8) &  67 &  1991 &  1 &  0.67102 &  0.67224(8) &  97 &  6136 &  2 &  0.55725 &  0.55797(2) &  127 &  13881 &  1 &  0.48406 &  0.48422(1) \\ 
 38 &  351 &  1 &  0.88879 &  0.88988(8) &  68 &  2087 &  2 &  0.66743 &  0.66920(6) &  98 &  6336 &  1 &  0.54944 &  0.54965(1) &  128 &  14222 &  2 &  0.48273 &  0.48305(3) \\ 
\end{tabular}

\end{ruledtabular}
\caption{\label{tab:E4}Comparison between the CF and the exact energies ($V_{\rm CF}$ and $V_{\rm ex}$) for
	$N=4$.  The symbols in this and the following tables have the following meaning:
	$L$ is the angular momentum; $D$ is the dimension of the full lowest LL Fock 
	space; and $D^*$ is the dimension of the CF basis.  The energies are in units of $e^2/\epsilon \ell$, where $\epsilon$ is the dielectric constant of the background 
	semiconductor and $\ell=\sqrt{\hbar c/eB}$ is the magnetic length.}
\end{table*}
\squeezetable
\begin{table*}
\begin{ruledtabular}
\begin{tabular}{r r r l l | r r r l l| r r r l l| r r r l l}
 $L$ &  $D$ & $D^*$ &  $V_{\rm ex}$ &  $V_{\rm CF}$ &  $L$ &  $D$ & $D^*$ &  $V_{\rm
 ex}$ &  $V_{\rm CF}$ &   $L$ &  $D$ & $D^*$ &  $V_{\rm ex}$ &  $V_{\rm CF}$ &  $L$ &
 $D$ & $D^*$ &  $V_{\rm ex}$ &  $V_{\rm CF}$ \\ 
\hline 
 13 &  3 &  1 &  3.20199 &  3.20205(29) &  39 &  603 &  4 &  1.80374 &  1.80580(23) &  65 &  5260 &  1 &  1.36535 &  1.36767(8) &  91 &  21224 &  1 &  1.15674 &  1.16090(6) \\ 
 14 &  5 &  1 &  3.05525 &  3.05557(34) &  40 &  674 &  1 &  1.75246 &  1.75282(18) &  66 &  5608 &  2 &  1.36535 &  1.36759(22) &  92 &  22204 &  2 &  1.15054 &  1.15487(9) \\ 
 15 &  7 &  1 &  2.91866 &  2.91868(43) &  41 &  748 &  3 &  1.75246 &  1.75288(31) &  67 &  5969 &  4 &  1.35517 &  1.35697(15) &  93 &  23212 &  4 &  1.14579 &  1.14689(10) \\ 
 16 &  10 &  3 &  2.91866 &  2.91927(60) &  42 &  831 &  1 &  1.72570 &  1.72659(20) &  68 &  6351 &  7 &  1.34694 &  1.34856(13) &  94 &  24260 &  2 &  1.14172 &  1.14284(19) \\ 
 17 &  13 &  1 &  2.83078 &  2.83119(66) &  43 &  918 &  2 &  1.71164 &  1.71548(21) &  69 &  6747 &  11 &  1.33969 &  1.34091(8) &  95 &  25337 &  1 &  1.12540 &  1.12654(8) \\ 
 18 &  18 &  1 &  2.69089 &  2.69274(9) &  44 &  1014 &  5 &  1.69696 &  1.70152(21) &  70 &  7166 &  1 &  1.31464 &  1.31980(9) &  96 &  26455 &  6 &  1.12540 &  1.12645(4) \\ 
 19 &  23 &  3 &  2.66332 &  2.66466(35) &  45 &  1115 &  1 &  1.64891 &  1.65141(11) &  71 &  7599 &  1 &  1.31464 &  1.31964(14) &  97 &  27604 &  2 &  1.11969 &  1.12238(3) \\ 
 20 &  30 &  1 &  2.53676 &  2.53706(42) &  46 &  1226 &  2 &  1.64894 &  1.65148(25) &  72 &  8056 &  2 &  1.30556 &  1.31141(4) &  98 &  28796 &  1 &  1.11535 &  1.12055(7) \\ 
 21 &  37 &  3 &  2.53676 &  2.53710(58) &  47 &  1342 &  4 &  1.63127 &  1.63345(10) &  73 &  8529 &  4 &  1.29822 &  1.29938(14) &  99 &  30020 &  4 &  1.11166 &  1.11286(10) \\ 
 22 &  47 &  1 &  2.42979 &  2.43092(41) &  48 &  1469 &  7 &  1.61635 &  1.61828(21) &  74 &  9027 &  2 &  1.29148 &  1.29258(16) &  100 &  31289 &  1 &  1.09648 &  1.09737(7) \\ 
 23 &  57 &  2 &  2.41248 &  2.41695(62) &  49 &  1602 &  11 &  1.60206 &  1.60277(31) &  75 &  9542 &  1 &  1.26920 &  1.27036(9) &  101 &  32591 &  3 &  1.09648 &  1.09738(7) \\ 
 24 &  70 &  5 &  2.37282 &  2.37632(85) &  50 &  1747 &  1 &  1.56144 &  1.56666(3) &  76 &  10083 &  6 &  1.26920 &  1.27029(12) &  102 &  33940 &  1 &  1.09120 &  1.09795(6) \\ 
 25 &  84 &  1 &  2.24724 &  2.24910(16) &  51 &  1898 &  1 &  1.56144 &  1.56655(18) &  77 &  10642 &  2 &  1.26100 &  1.26384(13) &  103 &  35324 &  2 &  1.08722 &  1.09060(9) \\ 
 26 &  101 &  2 &  2.23577 &  2.23796(46) &  52 &  2062 &  2 &  1.54661 &  1.55380(9) &  78 &  11229 &  1 &  1.25446 &  1.25857(11) &  104 &  36756 &  5 &  1.08384 &  1.08634(13) \\ 
 27 &  119 &  4 &  2.20889 &  2.21135(19) &  53 &  2233 &  4 &  1.53301 &  1.53401(14) &  79 &  11835 &  4 &  1.24894 &  1.25059(9) &  105 &  38225 &  1 &  1.06968 &  1.07123(3) \\ 
 28 &  141 &  7 &  2.17424 &  2.17563(68) &  54 &  2418 &  2 &  1.51746 &  1.51797(7) &  80 &  12470 &  1 &  1.22816 &  1.22917(4) &  106 &  39744 &  2 &  1.06968 &  1.07111(20) \\ 
 29 &  164 &  11 &  2.13794 &  2.13853(73) &  55 &  2611 &  1 &  1.48714 &  1.48776(10) &  81 &  13125 &  3 &  1.22816 &  1.22895(9) &  107 &  41301 &  4 &  1.06477 &  1.06622(4) \\ 
 30 &  192 &  1 &  2.02725 &  2.03083(17) &  56 &  2818 &  6 &  1.48714 &  1.48809(22) &  82 &  13811 &  1 &  1.22072 &  1.22684(12) &  108 &  42910 &  7 &  1.06110 &  1.06269(2) \\ 
 31 &  221 &  1 &  2.02725 &  2.03097(27) &  57 &  3034 &  2 &  1.47396 &  1.47634(15) &  83 &  14518 &  2 &  1.21493 &  1.21855(9) &  109 &  44559 &  11 &  1.05798 &  1.05946(3) \\ 
 32 &  255 &  2 &  1.99957 &  2.00534(11) &  58 &  3266 &  1 &  1.46155 &  1.46315(4) &  84 &  15257 &  5 &  1.21000 &  1.21295(24) &  110 &  46262 &  1 &  1.04475 &  1.04795(4) \\ 
 33 &  291 &  4 &  1.96593 &  1.96741(23) &  59 &  3507 &  4 &  1.45293 &  1.45490(12) &  85 &  16019 &  1 &  1.19085 &  1.19276(10) &  111 &  48006 &  1 &  1.04475 &  1.04789(4) \\ 
 34 &  333 &  2 &  1.92313 &  1.92424(17) &  60 &  3765 &  1 &  1.42240 &  1.42328(17) &  86 &  16814 &  2 &  1.19085 &  1.19276(3) &  112 &  49806 &  2 &  1.04018 &  1.04349(8) \\ 
 35 &  377 &  1 &  1.87634 &  1.87694(30) &  61 &  4033 &  3 &  1.42240 &  1.42339(17) &  87 &  17633 &  4 &  1.18408 &  1.18585(3) &  113 &  51649 &  4 &  1.03678 &  1.03755(7) \\ 
 36 &  427 &  6 &  1.87634 &  1.87662(29) &  62 &  4319 &  1 &  1.41061 &  1.41445(13) &  88 &  18487 &  7 &  1.17885 &  1.18048(10) &  114 &  53550 &  2 &  1.03389 &  1.03495(8) \\ 
 37 &  480 &  2 &  1.85037 &  1.85214(37) &  63 &  4616 &  2 &  1.40132 &  1.40514(12) &  89 &  19366 &  11 &  1.17438 &  1.17585(12) & \\ 
 38 &  540 &  1 &  1.81607 &  1.81678(9) &  64 &  4932 &  5 &  1.39332 &  1.39690(36) &  90 &  20282 &  1 &  1.15674 &  1.16095(10) & \\ 
\end{tabular}

\end{ruledtabular}
\caption{\label{tab:E5}Comparison between the CF and the exact energies ($V_{\rm CF}$ and $V_{\rm ex}$) for $N=5$.}
\end{table*}
\endgroup

\begingroup
\squeezetable
\begin{table*}
\begin{ruledtabular}
\begin{tabular}{r r r l l | r r r l l| r r r l l| r r r l l}
 $L$ &  $D$ & $D^*$ &  $V_{\rm ex}$ &  $V_{\rm CF}$ &  $L$ &  $D$ & $D^*$ &  $V_{\rm
 ex}$ &  $V_{\rm CF}$ &   $L$ &  $D$ & $D^*$ &  $V_{\rm ex}$ &  $V_{\rm CF}$ &  $L$ &
 $D$ & $D^*$ &  $V_{\rm ex}$ &  $V_{\rm CF}$ \\ 
\hline 
 19 &  5 &  1 &  4.52568 &  4.52563(84) &  52 &  2702 &  10 &  2.69635 &  2.70122(14) &  85 &  38677 &  1 &  2.06506 &  2.06929(9) &  118 &  216705 &  9 &  1.75766 &  1.76108(13) \\ 
 20 &  7 &  1 &  4.39138 &  4.39214(47) &  53 &  3009 &  5 &  2.66882 &  2.67239(64) &  86 &  41134 &  5 &  2.06506 &  2.06911(12) &  119 &  226479 &  3 &  1.74584 &  1.75185(20) \\ 
 21 &  11 &  1 &  4.26439 &  4.26485(31) &  54 &  3331 &  2 &  2.63071 &  2.63357(25) &  87 &  43752 &  2 &  2.05433 &  2.05522(17) &  120 &  236534 &  1 &  1.73124 &  1.73566(10) \\ 
 22 &  14 &  3 &  4.26439 &  4.26557(67) &  55 &  3692 &  1 &  2.58540 &  2.58872(13) &  88 &  46461 &  9 &  2.04622 &  2.04944(24) &  121 &  247010 &  3 &  1.73124 &  1.73575(13) \\ 
 23 &  20 &  2 &  4.15579 &  4.15623(25) &  56 &  4070 &  5 &  2.58541 &  2.58807(20) &  89 &  49342 &  3 &  2.02791 &  2.03308(20) &  122 &  257783 &  8 &  1.72727 &  1.73012(18) \\ 
 24 &  26 &  1 &  4.05541 &  4.05721(58) &  57 &  4494 &  2 &  2.55188 &  2.55252(31) &  90 &  52327 &  1 &  2.00538 &  2.00952(20) &  123 &  269005 &  2 &  1.72031 &  1.72323(11) \\ 
 25 &  35 &  1 &  3.92152 &  3.92355(10) &  58 &  4935 &  9 &  2.54880 &  2.55221(22) &  91 &  55491 &  3 &  2.00538 &  2.00969(35) &  124 &  280534 &  4 &  1.70935 &  1.71436(23) \\ 
 26 &  44 &  3 &  3.90771 &  3.90868(73) &  59 &  5427 &  3 &  2.51327 &  2.51647(68) &  92 &  58767 &  8 &  1.99893 &  2.00142(44) &  125 &  292534 &  1 &  1.69562 &  1.69987(6) \\ 
 27 &  58 &  2 &  3.79370 &  3.79420(72) &  60 &  5942 &  1 &  2.47124 &  2.47423(27) &  93 &  62239 &  2 &  1.98517 &  1.98615(10) &  126 &  304865 &  2 &  1.69562 &  1.69984(3) \\ 
 28 &  71 &  5 &  3.79370 &  3.79447(29) &  61 &  6510 &  3 &  2.47124 &  2.47420(61) &  94 &  65827 &  4 &  1.97151 &  1.97625(13) &  127 &  317683 &  5 &  1.69191 &  1.69581(16) \\ 
 29 &  90 &  2 &  3.69049 &  3.69200(92) &  62 &  7104 &  8 &  2.45835 &  2.45998(31) &  95 &  69624 &  1 &  1.95061 &  1.95495(22) &  128 &  330850 &  9 &  1.68552 &  1.68900(11) \\ 
 30 &  110 &  1 &  3.56719 &  3.56824(40) &  63 &  7760 &  2 &  2.42388 &  2.42431(35) &  96 &  73551 &  2 &  1.95061 &  1.95514(20) &  129 &  344534 &  1 &  1.67503 &  1.68120(11) \\ 
 31 &  136 &  3 &  3.56264 &  3.56580(66) &  64 &  8442 &  4 &  2.40947 &  2.41278(12) &  97 &  77695 &  5 &  1.94470 &  1.94832(11) &  130 &  358579 &  2 &  1.66210 &  1.66513(14) \\ 
 32 &  163 &  7 &  3.52932 &  3.53013(84) &  65 &  9192 &  1 &  2.37120 &  2.37547(20) &  98 &  81979 &  9 &  1.93471 &  1.93824(39) &  131 &  373165 &  4 &  1.66210 &  1.66519(27) \\ 
 33 &  199 &  2 &  3.41858 &  3.41952(30) &  66 &  9975 &  2 &  2.37120 &  2.37513(14) &  99 &  86499 &  1 &  1.91890 &  1.92278(10) &  132 &  388138 &  7 &  1.65863 &  1.66131(18) \\ 
 34 &  235 &  4 &  3.40210 &  3.40435(61) &  67 &  10829 &  5 &  2.35951 &  2.36237(52) &  100 &  91164 &  2 &  1.90010 &  1.90348(12) &  133 &  403670 &  12 &  1.65264 &  1.65522(33) \\ 
 35 &  282 &  1 &  3.28942 &  3.29204(19) &  68 &  11720 &  9 &  2.34182 &  2.34476(24) &  101 &  96079 &  4 &  1.90010 &  1.90329(29) &  134 &  419609 &  18 &  1.64273 &  1.64584(15) \\ 
 36 &  331 &  2 &  3.28587 &  3.29094(28) &  69 &  12692 &  1 &  2.30954 &  2.31154(17) &  102 &  101155 &  7 &  1.89472 &  1.89762(22) &  135 &  436140 &  1 &  1.63050 &  1.63876(7) \\ 
 37 &  391 &  5 &  3.25902 &  3.26178(31) &  70 &  13702 &  2 &  2.28245 &  2.28574(35) &  103 &  106491 &  12 &  1.88550 &  1.88831(53) &  136 &  453091 &  1 &  1.63050 &  1.63878(14) \\ 
 38 &  454 &  9 &  3.21604 &  3.21752(83) &  71 &  14800 &  4 &  2.28245 &  2.28624(68) &  104 &  111999 &  18 &  1.87118 &  1.87361(51) &  137 &  470660 &  2 &  1.62723 &  1.63454(7) \\ 
 39 &  532 &  1 &  3.11031 &  3.11221(29) &  72 &  15944 &  7 &  2.27239 &  2.27584(8) &  105 &  117788 &  1 &  1.85328 &  1.86170(17) &  138 &  488678 &  3 &  1.62159 &  1.62801(1) \\ 
 40 &  612 &  2 &  3.06846 &  3.07277(56) &  73 &  17180 &  12 &  2.25596 &  2.25924(17) &  106 &  123755 &  1 &  1.85328 &  1.86180(13) &  139 &  507334 &  4 &  1.61221 &  1.61833(7) \\ 
 41 &  709 &  4 &  3.06846 &  3.07263(46) &  74 &  18467 &  18 &  2.23266 &  2.23432(46) &  107 &  130019 &  2 &  1.84828 &  1.85551(23) &  140 &  526461 &  2 &  1.60064 &  1.60352(19) \\ 
 42 &  811 &  7 &  3.03681 &  3.03929(84) &  75 &  19858 &  1 &  2.20188 &  2.20932(19) &  108 &  136479 &  3 &  1.83963 &  1.84614(12) &  141 &  546261 &  1 &  1.60064 &  1.60808(6) \\ 
 43 &  931 &  12 &  3.00162 &  3.00288(91) &  76 &  21301 &  1 &  2.20188 &  2.20944(22) &  109 &  143247 &  4 &  1.82642 &  1.83222(16) &  142 &  566547 &  10 &  1.59756 &  1.60028(38) \\ 
 44 &  1057 &  18 &  2.95620 &  2.95640(61) &  77 &  22856 &  2 &  2.19230 &  2.19868(34) &  110 &  150224 &  2 &  1.80978 &  1.81276(9) &  143 &  587535 &  5 &  1.59225 &  1.59806(30) \\ 
 45 &  1206 &  1 &  2.86015 &  2.86444(21) &  78 &  24473 &  3 &  2.17671 &  2.18392(17) &  111 &  157532 &  1 &  1.80978 &  1.81507(10) &  144 &  609040 &  2 &  1.58332 &  1.58980(9) \\ 
 46 &  1360 &  1 &  2.86015 &  2.86427(33) &  79 &  26207 &  4 &  2.15698 &  2.16067(26) &  112 &  165056 &  10 &  1.80521 &  1.80836(23) &  145 &  631269 &  1 &  1.57236 &  1.57642(6) \\ 
 47 &  1540 &  2 &  2.83682 &  2.84249(21) &  80 &  28009 &  2 &  2.13038 &  2.13341(12) &  113 &  172929 &  5 &  1.79729 &  1.80306(20) &  146 &  654039 &  5 &  1.57236 &  1.57638(12) \\ 
 48 &  1729 &  3 &  2.80401 &  2.81080(37) &  81 &  29941 &  1 &  2.12855 &  2.13017(15) &  114 &  181038 &  2 &  1.78476 &  1.79071(14) &  147 &  677571 &  2 &  1.56945 &  1.57511(12) \\ 
 49 &  1945 &  4 &  2.76617 &  2.77264(33) &  82 &  31943 &  10 &  2.12260 &  2.12620(49) &  115 &  189509 &  1 &  1.76921 &  1.77344(27) &  148 &  701661 &  9 &  1.56444 &  1.56751(15) \\ 
 50 &  2172 &  2 &  2.71674 &  2.72205(36) &  83 &  34085 &  5 &
2.10896 &  2.11372(47) &  116 &  198230 &  5 &  1.76921 &  1.77351(18)
& \\
 51 &  2432 &  1 &  2.69635 &  2.70145(36) &  84 &  36308 &  2 &
2.08934 &  2.09405(13) &  117 &  207338 &  2 &  1.76490 &  1.76806(14)
& \\
\end{tabular}

\end{ruledtabular}
\caption{\label{tab:E6}Comparison between the CF and the exact energies ($V_{\rm CF}$ and $V_{\rm ex}$) for $N=6$.}
\end{table*}
\endgroup

The projection into the lowest LL 
is accomplished by the method outlined in the
literature.~\cite{Kamilla}  To give a simple example, 
consider the state $[N-1,1]$ state in Eq.~(\ref{wf2}).
We use the identity (apart from the constant factor),
\begin{equation}
A\left[ z_{1}^*\cdot\prod_{i=2}^{N}  z_i^{i-2} \right]
\prod_{j<k} (z_j-z_k)^{2p}
= A\left[ z_{1}^* J_1^p \cdot\prod_{i=2}^{N}  (z_i^{i-2} J_i^p) \right]
\end{equation}
with
\begin{equation}
J_i \equiv \prod_{j(\ne i)} (z_i - z_j)
\end{equation}
and project each element to the lowest LL, resulting in
\begin{equation}
\Psi =  e^{-\sum_{l=1}^N |z_l|^2/4}  A\left[ {\cal P}_{\rm LLL}(z_{1}^*J_1^p)\cdot
\prod_{i=2}^{N} {\cal P}_{\rm LLL}(z_i^{i-2} J_i^p)\right].
\end{equation}
The final step is achieved by 
placing all of the $z_j^*$'s to the
left, and substituting $2\partial/\partial z_j$ into $z_j^*$ with the
assumption that the derivatives do not act on the Gaussian
factor.~\cite{Jain,Girvin2}
The resulting wave function is then given by
\begin{equation}
\Psi =  A\left[ \sum_{j\ne 1}\frac{1}{z_1-z_j}\cdot
\prod_{i=2}^{N} z_i^{i-2} \right]
\Phi_1^{2p}.
\end{equation}
Lowest LL projection of other can be accomplished similarly, although 
the details are more complicated.

\begingroup
\squeezetable
\begin{table*}
\begin{ruledtabular}
\begin{tabular}{r r r l l | r r r l l| r r r l l| r r r l l}
 $L$ &  $D$ & $D^*$ &  $V_{\rm ex}$ &  $V_{\rm CF}$ &  $L$ &  $D$ & $D^*$ &  $V_{\rm
 ex}$ &  $V_{\rm CF}$ &   $L$ &  $D$ & $D^*$ &  $V_{\rm ex}$ &  $V_{\rm CF}$ &  $L$ &
 $D$ & $D^*$ &  $V_{\rm ex}$ &  $V_{\rm CF}$ \\ 
\hline 
 26 &  7 &  1 &  6.04656 &  6.04681(54) &  52 &  2093 &  2 &  4.26158 &  4.26561(73) &  78 &  34082 &  1 &  3.42977 &  3.43487(25) &  104 &  225286 &  29 &  2.94535 &  2.94866(61) \\ 
 27 &  11 &  1 &  5.92221 &  5.92332(86) &  53 &  2400 &  5 &  4.23491 &  4.23965(35) &  79 &  37108 &  9 &  3.41304 &  3.41884(47) &  105 &  239691 &  1 &  2.91436 &  2.93077(20) \\ 
 28 &  15 &  1 &  5.80240 &  5.80314(97) &  54 &  2738 &  9 &  4.19836 &  4.20134(65) &  80 &  40340 &  3 &  3.37769 &  3.38473(36) &  106 &  254826 &  1 &  2.91436 &  2.93053(15) \\ 
 29 &  21 &  4 &  5.80240 &  5.80278(98) &  55 &  3120 &  17 &  4.16124 &  4.16382(75) &  81 &  43819 &  1 &  3.33410 &  3.33890(34) &  107 &  270775 &  2 &  2.90600 &  2.92167(22) \\ 
 30 &  28 &  2 &  5.70400 &  5.70370(82) &  56 &  3539 &  1 &  4.07199 &  4.07643(27) &  82 &  47527 &  5 &  3.33410 &  3.33892(27) &  108 &  287521 &  3 &  2.89354 &  2.90824(20) \\ 
 31 &  38 &  2 &  5.56817 &  5.56805(59) &  57 &  4011 &  2 &  4.01867 &  4.02474(63) &  83 &  51508 &  1 &  3.32037 &  3.32853(26) &  109 &  305146 &  5 &  2.88096 &  2.89366(14) \\ 
 32 &  49 &  1 &  5.47914 &  5.48224(86) &  58 &  4526 &  4 &  4.01774 &  4.02173(71) &  84 &  55748 &  5 &  3.30206 &  3.30849(32) &  110 &  323633 &  4 &  2.86002 &  2.86857(82) \\ 
 33 &  65 &  1 &  5.35609 &  5.35854(66) &  59 &  5102 &  7 &  3.99569 &  4.00145(65) &  85 &  60289 &  1 &  3.28254 &  3.29191(18) &  111 &  343074 &  2 &  2.83206 &  2.83653(21) \\ 
 34 &  82 &  4 &  5.35422 &  5.35502(75) &  60 &  5731 &  12 &  3.96390 &  3.96918(64) &  86 &  65117 &  5 &  3.25276 &  3.26002(32) &  112 &  363446 &  1 &  2.83206 &  2.85264(22) \\ 
 35 &  105 &  2 &  5.24786 &  5.24862(35) &  61 &  6430 &  19 &  3.92950 &  3.93103(66) &  87 &  70281 &  1 &  3.21251 &  3.21788(44) &  113 &  384845 &  18 &  2.82445 &  2.82965(75) \\ 
 36 &  131 &  1 &  5.15820 &  5.16479(47) &  62 &  7190 &  29 &  3.88434 &  3.88684(56) &  88 &  75762 &  3 &  3.21251 &  3.21794(49) &  114 &  407254 &  9 &  2.81316 &  2.82007(21) \\ 
 37 &  164 &  5 &  5.14271 &  5.14442(88) &  63 &  8033 &  1 &  3.79495 &  3.80220(41) &  89 &  81612 &  8 &  3.20117 &  3.20516(23) &  115 &  430768 &  5 &  2.80176 &  2.80944(22) \\ 
 38 &  201 &  2 &  5.03005 &  5.03343(78) &  64 &  8946 &  1 &  3.79495 &  3.80267(56) &  90 &  87816 &  1 &  3.18371 &  3.19566(34) &  116 &  455370 &  2 &  2.78234 &  2.79337(13) \\ 
 39 &  248 &  1 &  4.91568 &  4.91829(53) &  65 &  9953 &  2 &  3.77448 &  3.78136(54) &  91 &  94425 &  3 &  3.16614 &  3.17410(51) &  117 &  481165 &  1 &  2.75630 &  2.76250(23) \\ 
 40 &  300 &  4 &  4.91568 &  4.91851(73) &  66 &  11044 &  3 &  3.74470 &  3.75302(52) &  92 &  101423 &  7 &  3.14026 &  3.14628(52) &  118 &  508130 &  8 &  2.75630 &  2.76300(33) \\ 
 41 &  364 &  1 &  4.83425 &  4.84080(65) &  67 &  12241 &  5 &  3.71216 &  3.71858(57) &  93 &  108869 &  1 &  3.10327 &  3.10994(8) &  119 &  536375 &  3 &  2.75630 &  2.76198(87) \\ 
 42 &  436 &  4 &  4.79274 &  4.79393(72) &  68 &  13534 &  4 &  3.67128 &  3.68989(18) &  94 &  116742 &  2 &  3.10327 &  3.11018(33) &  120 &  565883 &  1 &  2.73889 &  2.75099(12) \\ 
 43 &  522 &  1 &  4.72350 &  4.72758(51) &  69 &  14950 &  2 &  3.62471 &  3.63773(40) &  95 &  125104 &  5 &  3.09306 &  3.10009(18) &  121 &  596763 &  9 &  2.72857 &  2.73428(26) \\ 
 44 &  618 &  4 &  4.66883 &  4.67236(63) &  70 &  16475 &  1 &  3.61998 &  3.63879(45) &  96 &  133939 &  9 &  3.07797 &  3.08298(37) &  122 &  628998 &  3 &  2.71052 &  2.72085(53) \\ 
 45 &  733 &  1 &  4.55743 &  4.55945(83) &  71 &  18138 &  18 &  3.60749 &  3.61584(47) &  97 &  143307 &  17 &  3.06295 &  3.06674(32) &  123 &  662708 &  1 &  2.68631 &  2.69274(14) \\ 
 46 &  860 &  3 &  4.55743 &  4.55947(42) &  72 &  19928 &  9 &  3.58162 &  3.58962(41) &  98 &  153192 &  1 &  3.03687 &  3.04788(18) &  124 &  697870 &  5 &  2.68631 &  2.69278(30) \\ 
 47 &  1009 &  8 &  4.52734 &  4.52891(39) &  73 &  21873 &  5 &  3.55098 &  3.55586(43) &  99 &  163662 &  2 &  3.00453 &  3.01093(18) &  125 &  734609 &  1 &  2.67986 &  2.69490(26) \\ 
 48 &  1175 &  1 &  4.45781 &  4.46275(22) &  74 &  23961 &  2 &  3.51776 &  3.52660(29) &  100 &  174696 &  4 &  3.00453 &  3.01048(13) &  126 &  772909 &  5 &  2.67032 &  2.67658(14) \\ 
 49 &  1367 &  3 &  4.40435 &  4.40694(84) &  75 &  26226 &  1 &  3.47044 &  3.47617(10) &  101 &  186366 &  7 &  2.99539 &  3.00227(34) &  127 &  812893 &  1 &  2.66074 &  2.67623(21) \\ 
 50 &  1579 &  7 &  4.36566 &  4.36983(89) &  76 &  28652 &  8 &  3.47044 &  3.47749(52) &  102 &  198655 &  12 &  2.98187 &  2.98934(91) & \\ 
 51 &  1824 &  1 &  4.26158 &  4.26604(23) &  77 &  31275 &  3 &  3.45496 &  3.46606(50) &  103 &  211634 &  19 &  2.96834 &  2.97506(53) & \\ 
\end{tabular}

\end{ruledtabular}
\caption{\label{tab:E7}Comparison between the CF and the exact energies ($V_{\rm CF}$ and $V_{\rm ex}$) for $N=7$.}
\end{table*}
\endgroup
\begingroup
\squeezetable
\begin{table*}
\begin{ruledtabular}
\begin{tabular}{r r r l l | r r r l l| r r r l l| r r r l l}
 $L$ &  $D$ & $D^*$ &  $V_{\rm ex}$ &  $V_{\rm CF}$ &  $L$ &  $D$ & $D^*$ &  $V_{\rm
 ex}$ &  $V_{\rm CF}$ &   $L$ &  $D$ & $D^*$ &  $V_{\rm ex}$ &  $V_{\rm CF}$ &  $L$ &
 $D$ & $D^*$ &  $V_{\rm ex}$ &  $V_{\rm CF}$ \\ 
\hline 
 33 &  7 &  1 &  7.8710 &  7.8706(11) &  55 &  1527 &  5 &  6.1263 &  6.1282(9) &  77 &  27493 &  2 &  5.0769 &  5.0829(6) &  99 &  207945 &  13 &  4.4669 &  4.4765(4) \\ 
 34 &  11 &  2 &  7.7485 &  7.7482(5) &  56 &  1801 &  2 &  6.0235 &  6.0251(7) &  78 &  30588 &  4 &  5.0603 &  5.0647(7) &  100 &  225132 &  6 &  4.4502 &  4.4614(11) \\ 
 35 &  15 &  2 &  7.6316 &  7.6318(20) &  57 &  2104 &  6 &  6.0235 &  6.0263(10) &  79 &  33940 &  7 &  5.0527 &  5.0579(9) &  101 &  243434 &  2 &  4.4221 &  4.4320(4) \\ 
 36 &  22 &  1 &  7.5176 &  7.5191(4) &  58 &  2462 &  1 &  5.9294 &  5.9380(7) &  80 &  37638 &  12 &  5.0190 &  5.0266(8) &  102 &  263081 &  1 &  4.3873 &  4.3940(2) \\ 
 37 &  29 &  5 &  7.5176 &  7.5181(11) &  59 &  2857 &  4 &  5.9212 &  5.9292(6) &  81 &  41635 &  19 &  4.9876 &  4.9932(2) &  103 &  283981 &  10 &  4.3859 &  4.3881(5) \\ 
 38 &  40 &  3 &  7.3979 &  7.3977(3) &  60 &  3319 &  1 &  5.8259 &  5.8282(7) &  82 &  46031 &  30 &  4.9521 &  4.9567(4) &  104 &  306376 &  4 &  4.3495 &  4.3559(3) \\ 
 39 &  52 &  3 &  7.2975 &  7.2983(15) &  61 &  3828 &  3 &  5.8116 &  5.8182(9) &  83 &  50774 &  44 &  4.9183 &  4.9215(12) &  105 &  330170 &  1 &  4.3104 &  4.3168(5) \\ 
 40 &  70 &  2 &  7.1704 &  7.1711(6) &  62 &  4417 &  9 &  5.7565 &  5.7612(19) &  84 &  55974 &  1 &  4.8299 &  4.8373(6) &  106 &  355626 &  10 &  4.3104 &  4.3161(9) \\ 
 41 &  89 &  1 &  7.0899 &  7.0921(10) &  63 &  5066 &  1 &  5.6543 &  5.6569(6) &  85 &  61575 &  1 &  4.8299 &  4.8376(5) &  107 &  382641 &  3 &  4.2946 &  4.3018(7) \\ 
 42 &  116 &  1 &  6.9766 &  6.9786(8) &  64 &  5812 &  4 &  5.6422 &  5.6479(2) &  86 &  67696 &  2 &  4.8101 &  4.8178(4) &  108 &  411498 &  1 &  4.2604 &  4.2654(6) \\ 
 43 &  146 &  5 &  6.9723 &  6.9735(23) &  65 &  6630 &  9 &  5.6246 &  5.6273(6) &  87 &  74280 &  3 &  4.7816 &  4.7910(7) &  109 &  442089 &  5 &  4.2521 &  4.2565(3) \\ 
 44 &  186 &  3 &  6.8818 &  6.8830(5) &  66 &  7564 &  1 &  5.5388 &  5.5427(4) &  88 &  81457 &  5 &  4.7501 &  4.7589(6) &  110 &  474715 &  1 &  4.2291 &  4.2325(1) \\ 
 45 &  230 &  1 &  6.7912 &  6.7987(5) &  67 &  8588 &  2 &  5.5243 &  5.5339(7) &  89 &  89162 &  7 &  4.7173 &  4.7603(4) &  111 &  509267 &  7 &  4.2071 &  4.2144(8) \\ 
 46 &  288 &  1 &  6.6696 &  6.6778(5) &  68 &  9749 &  6 &  5.4806 &  5.4845(6) &  90 &  97539 &  4 &  4.6754 &  4.7122(3) &  112 &  546067 &  2 &  4.1686 &  4.1735(1) \\ 
 47 &  352 &  5 &  6.6648 &  6.6673(10) &  69 &  11018 &  12 &  5.4370 &  5.4434(15) &  91 &  106522 &  2 &  4.6418 &  4.6674(5) &  113 &  584996 &  7 &  4.1686 &  4.1736(6) \\ 
 48 &  434 &  3 &  6.5610 &  6.5627(10) &  70 &  12450 &  1 &  5.3388 &  5.3429(1) &  92 &  116263 &  1 &  4.6384 &  4.6787(3) &  114 &  626401 &  1 &  4.1463 &  4.1508(4) \\ 
 49 &  525 &  1 &  6.4566 &  6.4594(6) &  71 &  14012 &  2 &  5.3322 &  5.3354(9) &  93 &  126692 &  29 &  4.6235 &  4.6430(18) &  115 &  670162 &  5 &  4.1342 &  4.1390(4) \\ 
 50 &  638 &  6 &  6.4566 &  6.4590(6) &  72 &  15765 &  5 &  5.2966 &  5.2998(6) &  94 &  137977 &  17 &  4.5936 &  4.6101(20) &  116 &  716644 &  1 &  4.1112 &  4.1148(4) \\ 
 51 &  764 &  2 &  6.3745 &  6.3811(29) &  73 &  17674 &  9 &  5.2755 &  5.2802(6) &  95 &  150042 &  9 &  4.5644 &  4.5787(4) &  117 &  765722 &  3 &  4.1022 &  4.1129(3) \\ 
 52 &  919 &  1 &  6.2652 &  6.2738(5) &  74 &  19805 &  17 &  5.2422 &  5.2470(10) &  96 &  163069 &  5 &  4.5277 &  4.5395(4) &  118 &  817789 &  10 &  4.0758 &  4.0840(9) \\ 
 53 &  1090 &  4 &  6.2635 &  6.2713(7) &  75 &  22122 &  28 &  5.2101 &  5.2135(24) &  97 &  176978 &  2 &  4.5089 &  4.5254(3) & \\ 
 54 &  1297 &  1 &  6.1683 &  6.1722(6) &  76 &  24699 &  1 &  5.1250 &  5.1323(4) &  98 &  191964 &  1 &  4.4669 &  4.4770(4) & \\ 
\end{tabular}

\end{ruledtabular}
\caption{\label{tab:E8}Comparison between the CF and the exact energies ($V_{\rm CF}$ and $V_{\rm ex}$) for $N=8$.}
\end{table*}
\squeezetable
\begin{table*}
\begin{ruledtabular}
\begin{tabular}{r r r l l | r r r l l| r r r l l| r r r l l}
 $L$ &  $D$ & $D^*$ &  $V_{\rm ex}$ &  $V_{\rm CF}$ &  $L$ &  $D$ & $D^*$ &  $V_{\rm
 ex}$ &  $V_{\rm C\rm CF}$ &   $L$ &  $D$ & $D^*$ &  $V_{\rm ex}$ &  $V_{\rm CF}$ &  $L$ &
 $D$ & $D^*$ &  $V_{\rm ex}$ &  $V_{\rm CF}$ \\ 
\hline 
 42 &  11 &  1 &  9.7329 &  9.7325(16) &  61 &  1291 &  10 &  8.1734 &  8.1758(20) &  80 &  22380 &  1 &  7.0217 &  7.0237(3) &  99 &  177884 &  1 &  6.2652 &  6.2749(5) \\ 
 43 &  15 &  2 &  9.6181 &  9.6176(9) &  62 &  1549 &  4 &  8.0839 &  8.0924(14) &  81 &  25331 &  3 &  7.0110 &  7.0173(6) &  100 &  195666 &  2 &  6.2306 &  6.2352(9) \\ 
 44 &  22 &  2 &  9.5068 &  9.5088(20) &  63 &  1845 &  2 &  7.9840 &  7.9933(6) &  82 &  28629 &  8 &  6.9928 &  6.9962(19) &  101 &  214944 &  4 &  6.1974 &  6.2042(7) \\ 
 45 &  30 &  1 &  9.3978 &  9.3983(10) &  64 &  2194 &  1 &  7.8813 &  7.8912(9) &  83 &  32278 &  19 &  6.9402 &  6.9449(26) &  102 &  235899 &  7 &  6.1911 &  6.1957(16) \\ 
 46 &  41 &  15 &  9.3900 &  9.3894(35) &  65 &  2592 &  5 &  7.8813 &  7.8904(5) &  84 &  36347 &  2 &  6.8463 &  6.8500(9) &  103 &  258569 &  12 &  6.1689 &  6.1762(13) \\ 
 47 &  54 &  5 &  9.2860 &  9.2870(10) &  66 &  3060 &  2 &  7.7980 &  7.8030(13) &  85 &  40831 &  5 &  6.8151 &  6.8225(10) &  104 &  283161 &  19 &  6.1362 &  6.1422(20) \\ 
 48 &  73 &  6 &  9.1587 &  9.1578(13) &  67 &  3589 &  9 &  7.7651 &  7.7676(24) &  86 &  45812 &  12 &  6.8000 &  6.8060(7) &  105 &  309729 &  30 &  6.1025 &  6.1092(12) \\ 
 49 &  94 &  2 &  9.0657 &  9.0662(5) &  68 &  4206 &  3 &  7.6707 &  7.6748(4) &  87 &  51294 &  1 &  6.7133 &  6.7169(9) &  106 &  338484 &  45 &  6.0679 &  6.0749(6) \\ 
 50 &  123 &  2 &  8.9478 &  8.9482(10) &  69 &  4904 &  1 &  7.6174 &  7.6263(8) &  88 &  57358 &  2 &  6.7060 &  6.7125(7) &  107 &  369499 &  66 &  6.0451 &  6.0493(29) \\ 
 51 &  157 &  1 &  8.8745 &  8.8778(9) &  70 &  5708 &  5 &  7.5741 &  7.5837(10) &  89 &  64015 &  5 &  6.6849 &  6.6906(10) &  108 &  403016 &  1 &  5.9559 &  5.9635(7) \\ 
 52 &  201 &  1 &  8.7686 &  8.7699(9) &  71 &  6615 &  1 &  7.4741 &  7.4850(7) &  90 &  71362 &  11 &  6.6427 &  6.6500(6) &  109 &  439100 &  1 &  5.9559 &  5.9635(3) \\ 
 53 &  252 &  9 &  8.7625 &  8.7635(27) &  72 &  7657 &  6 &  7.4729 &  7.4817(9) &  91 &  79403 &  21 &  6.6008 &  6.6069(8) &  110 &  478025 &  2 &  5.9366 &  5.9452(7) \\ 
 54 &  318 &  5 &  8.6394 &  8.6383(85) &  73 &  8824 &  1 &  7.3809 &  7.3839(10) &  92 &  88252 &  1 &  6.5086 &  6.5136(6) &  111 &  519880 &  3 &  5.9107 &  5.9195(10) \\ 
 55 &  393 &  3 &  8.5800 &  8.5856(15) &  74 &  10156 &  5 &  7.3711 &  7.3779(12) &  93 &  97922 &  2 &  6.4891 &  6.4940(9) &  112 &  564945 &  5 &  5.8797 &  5.8894(6) \\ 
 56 &  488 &  1 &  8.4762 &  8.4857(5) &  75 &  11648 &  1 &  7.2951 &  7.3010(7) &  94 &  108527 &  5 &  6.4504 &  6.4554(13) &  113 &  613331 &  7 &  5.8477 &  5.8555(12) \\ 
 57 &  598 &  1 &  8.3625 &  8.3713(4) &  76 &  13338 &  4 &  7.2324 &  7.2357(8) &  95 &  120092 &  9 &  6.4416 &  6.4462(6) &  114 &  665355 &  7 &  5.8130 &  5.8867(2) \\ 
 58 &  732 &  6 &  8.3625 &  8.3691(22) &  77 &  15224 &  12 &  7.2148 &  7.2222(18) &  96 &  132751 &  17 &  6.4124 &  6.4172(13) &  115 &  721125 &  4 &  5.7746 &  5.8360(3) \\ 
 59 &  887 &  3 &  8.2705 &  8.2739(11) &  78 &  17354 &  2 &  7.1225 &  7.1320(7) &  97 &  146520 &  28 &  6.3827 &  6.3863(10) &  116 &  780997 &  2 &  5.7569 &  5.8087(2) \\ 
 60 &  1076 &  2 &  8.1734 &  8.1761(8) &  79 &  19720 &  6 &  7.1176 &  7.1253(14) &  98 &  161554 &  47 &  6.3501 &  6.3536(8) & \\ 
\end{tabular}

\end{ruledtabular}
\caption{\label{tab:E9}Comparison between the CF and the exact energies ($V_{\rm CF}$ and $V_{\rm ex}$) for $N=9$.}
\end{table*}
\squeezetable
\begin{table*}
\begin{ruledtabular}
\begin{tabular}{r r r l l | r r r l l| r r r l l| r r r l l}
 $L$ &  $D$ & $D^*$ &  $V_{\rm ex}$ &  $V_{\rm CF}$ &  $L$ &  $D$ & $D^*$ &  $V_{\rm
 ex}$ &  $V_{\rm CF}$ &   $L$ &  $D$ & $D^*$ &  $V_{\rm ex}$ &  $V_{\rm CF}$ &  $L$ &
 $D$ & $D^*$ &  $V_{\rm ex}$ &  $V_{\rm CF}$ \\ 
\hline 
 50 &  7 &  6 &  11.9863 &  11.9860(20) &  66 &  653 &  6 &  10.4788 &  10.4811(27) &  82 &  10936 &  2 &  9.4127 &  9.4293(6) &  98 &  89623 &  3 &  8.5745 &  8.5843(18) \\ 
 51 &  11 &  5 &  11.8667 &  11.8669(24) &  67 &  807 &  2 &  10.4256 &  10.4327(6) &  83 &  12690 &  1 &  9.3202 &  9.3308(9) &  99 &  100654 &  11 &  8.5278 &  8.5317(24) \\ 
 52 &  15 &  4 &  11.7534 &  11.7551(12) &  68 &  984 &  1 &  10.3294 &  10.3375(11) &  84 &  14663 &  6 &  9.2887 &  9.2982(17) &  100 &  112804 &  1 &  8.4802 &  8.4853(5) \\ 
 53 &  22 &  3 &  11.6444 &  11.6458(8) &  69 &  1204 &  1 &  10.2240 &  10.2332(21) &  85 &  16928 &  2 &  9.1955 &  9.2056(12) &  101 &  126299 &  4 &  8.4092 &  8.4163(14) \\ 
 54 &  30 &  1 &  11.5378 &  11.5388(15) &  70 &  1455 &  9 &  10.2240 &  10.2312(42) &  86 &  19466 &  9 &  9.1955 &  9.2052(12) &  102 &  141136 &  11 &  8.3965 &  8.4038(17) \\ 
 55 &  42 &  1 &  11.4332 &  11.4328(8) &  71 &  1761 &  4 &  10.1458 &  10.1473(16) &  87 &  22367 &  3 &  9.1116 &  9.1171(28) &  103 &  157564 &  1 &  8.3072 &  8.3110(3) \\ 
 56 &  55 &  18 &  11.4270 &  11.4273(33) &  72 &  2112 &  2 &  10.0508 &  10.0564(6) &  88 &  25608 &  12 &  9.1038 &  9.1127(12) &  104 &  175586 &  3 &  8.2823 &  8.2898(5) \\ 
 57 &  75 &  18 &  11.3010 &  11.3010(9) &  73 &  2534 &  1 &  9.9478 &  9.9625(15) &  89 &  29292 &  3 &  9.0219 &  9.0330(18) &  105 &  195491 &  9 &  8.2716 &  8.2805(5) \\ 
 58 &  97 &  11 &  11.2044 &  11.2052(9) &  74 &  3015 &  9 &  9.9478 &  9.9545(12) &  90 &  33401 &  1 &  8.9264 &  8.9346(11) &  106 &  217280 &  19 &  8.2485 &  8.2536(8) \\ 
 59 &  128 &  13 &  11.0855 &  11.0864(15) &  75 &  3590 &  5 &  9.8563 &  9.8679(14) &  91 &  38047 &  4 &  8.9148 &  8.9258(17) &  107 &  241279 &  1 &  8.1759 &  8.1818(7) \\ 
 60 &  164 &  6 &  10.9979 &  11.0002(14) &  76 &  4242 &  2 &  9.7660 &  9.7782(17) &  92 &  43214 &  14 &  8.8712 &  8.8822(12) &  108 &  267507 &  3 &  8.1204 &  8.1261(20) \\ 
 61 &  212 &  5 &  10.8879 &  10.8895(18) &  77 &  5013 &  1 &  9.6703 &  9.6807(8) &  93 &  49037 &  3 &  8.7784 &  8.7886(8) &  109 &  296320 &  8 &  8.0824 &  8.0906(16) \\ 
 62 &  267 &  1 &  10.8212 &  10.8254(27) &  78 &  5888 &  7 &  9.6703 &  9.6803(15) &  94 &  55494 &  10 &  8.7710 &  8.7808(10) &  110 &  327748 &  17 &  8.0589 &  8.0661(7) \\ 
 63 &  340 &  1 &  10.7207 &  10.7232(13) &  79 &  6912 &  3 &  9.5992 &  9.6085(26) &  95 &  62740 &  2 &  8.6840 &  8.6886(6) &  111 &  362198 &  1 &  7.9774 &  7.9822(8) \\ 
 64 &  423 &  19 &  10.7050 &  10.7051(36) &  80 &  8070 &  1 &  9.5034 &  9.5225(7) &  96 &  70760 &  7 &  8.6704 &  8.6790(12) &  112 &  399705 &  2 &  7.9717 &  7.9757(2) \\ 
 65 &  530 &  11 &  10.5949 &  10.5976(40) &  81 &  9418 &  8 &  9.4836 &  9.4884(21) &  97 &  79725 &  1 &  8.5839 &  8.5912(7) &  113 &  440725 &  5 &  7.9457 &  7.9511(4) \\ 
\end{tabular}

\end{ruledtabular}
\caption{\label{tab:E10}Comparison between the CF and the exact energies ($V_{\rm CF}$ and $V_{\rm ex}$) for $N=10$.}
\end{table*}
\endgroup

\section{Numerical methods}

\subsection{Exact diagonalization}

We calculate the exact interaction energy for small systems 
by either numerical diagonalization
using standard routines for small $L$, or a modified Lanczos algorithm for larger $L$.
In either cases, it is essential to know the Coulomb matrix elements.
In the second quantization language, the Coulomb Hamiltonian is written as
\begin{equation}
   {\cal H} = \frac 1 2\sum_{r,s,t,u}\,\langle r,s|V|t,u\rangle\,
              a_r^\dagger\,a_s^\dagger\,a_u\,a_t,
\end{equation}
where the operator $a_r^\dagger$ ($a_r$) creates (annihilates) an particle at state $|r\rangle$ with
angular momentum $r$.
$V$ is the Coulomb interaction in real space and $\langle r,s|V|t,u\rangle$ is its restriction
to the lowest Landau level Hilbert space.
Several expressions for $\langle r,s|V|t,u\rangle$ exist in the literature.~\cite{Girvin,Dev,Stone}
In this work we use the formula derived by Tsiper~\cite{Tsiper}:

\begin{widetext}
\begin{equation}
    \langle s+r, t|V|s,t+r\rangle =
    \sqrt{\frac{(s+r)!(t+r)!}{s!t!}}\,\frac{\Gamma(r+s+t+\frac 3 2)}{\pi\,2^{r+s+t+2}}
    [A_{st}^r\,B_{ts}^r+B_{st}^r\,A_{ts}^r], \label{Tsiper}
\end{equation}
where
\begin{equation}
   A_{st}^r =\sum_{i=0}^s \left(\begin{array}{c} s \\ i \end{array}\right)
              \frac{\Gamma(\frac 1 2 +i)\Gamma(\frac 1 2 + r+i)}
                   {(r+i)!\Gamma(\frac 3 2 +r+t+i)},\,\,\,\,\,\,\,\,\,\,
   B_{st}^r =\sum_{i=0}^s \left(\begin{array}{c} s \\ i \end{array}\right)
              \frac{\Gamma(\frac 1 2 +i)\Gamma(\frac 1 2 + r+i)}
                   {(r+i)!\Gamma(\frac 3 2 +r+t+i)}(\frac 1 2 +r+2i). \label{summation}
\end{equation}
\end{widetext}
Each term in the sum on the right hand side of Eq. (\ref{summation})
is positive definite, which makes this expression more stable in numerical 
calculations, which is especially important for large $L$.

Exact numerical diagonalization is limited to
systems with small numbers of particles at small angular momentum $L$,
 because the size of
the Hilbert space grows exponentially fast with $N$ and $L$.
For example, for $N=6$, the Fock space dimension grows from 21 to 701661 as 
$L$ is increased from 21 to 148.
When $D$ becomes large, we have obtained the ground state energy 
using the modified Lanczos algorithm of Gagliano {\em et al.}.~\cite{Gagliano}
Briefly, the algorithm begins with an initial guess of the ground state $|\psi_0\rangle$.
By applying the Hamiltonian ${\cal H}$ onto $|\psi_0\rangle$,
a new state $|\psi_1\rangle$ is defined as the following
\begin{equation}
   |\psi_1\rangle = \frac{{\cal H}|\psi_0\rangle-\langle{\cal H}\rangle|\psi_0\rangle}
                          {\sqrt{\langle{\cal H}^2\rangle-\langle{\cal H}\rangle^2}},
\end{equation}
where the notation $\langle{\cal H}^n\rangle$ represents the expectation value
$\langle\psi_0|{\cal H}^n|\psi_0\rangle$. It is straightforward to verify that $|\psi_1\rangle$ is
normalized and orthogonal to $|\psi_0\rangle$ by this construction.
The Hamiltonian ${\cal H}$ is now diagonalized 
in the subspace spanned by $\{|\psi_0\rangle,\,|\psi_1\rangle\}$, and 
the lowest eigenvalue $\varepsilon$ and its corresponding
eigenvector $|\tilde{\psi_1}\rangle$ are chosen 
as a better approximation the to true ground state and its energy
than $\langle{\cal H}\rangle$ and $|\psi_0\rangle$.
In terms of $\langle{\cal H}^n\rangle$ and $|\psi_0\rangle$, the ``better'' ground state energy
and corresponding state can be written as
\begin{eqnarray}
   \varepsilon &= & \langle{\cal H}\rangle+b\alpha,\\
   |\tilde{\psi_0}\rangle &=&\frac{|\psi_0\rangle+\alpha|\psi_1\rangle}{\sqrt{1+\alpha^2}},
\end{eqnarray}
in which
\begin{eqnarray}
   \alpha &=&f-\sqrt{f^2+1},\\
   f &=&\frac{\langle{\cal H}^3\rangle-3\langle{\cal H}\rangle\langle{\cal H}^2\rangle+2\langle{\cal H}\rangle^3}
            {2(\langle{\cal H}^2\rangle-\langle{\cal H}\rangle^2)^{3/2}},\\
   b &=&\sqrt{\langle{\cal H}^2\rangle-\langle{\cal H}\rangle^2}.
\end{eqnarray}
The state $|\tilde{\psi_0}\rangle$ is then used as a new initial guess and
the entire procedure is iterated until the relative energy difference
$|\varepsilon-\langle{\cal H}\rangle|$ is smaller than a predefined tolerance.

We place the program on a node of a Beowulf type PC cluster, each node consists a
dual PentiumIII 1GHz processor. The modified Lanczos algorithm converges
relatively fast to the true ground state. The most time consuming part of the calculation is
the Coulomb matrix element, especially when $L$ is very large. For example, when $N=6$ it takes
approximately 24 CPU hours to obtain the ground state energy for $L=135$.  
The largest system we have studied by the Lanczos method 
has a Fock space dimension of 817789.
We note that for large $L$, our energies are slightly lower than those in 
Ref.~\onlinecite{Landman},
presumably because they work with a truncated basis.

\subsection{CF diagonalization \label{subsec:CFdiag}}

CF diagonalization refers to a diagonalization of the full Coulomb Hamiltonian 
in a correlated CF basis.  The basis wave functions correspond to states with 
low ``CF kinetic energies"  (the kinetic energy is given by $\sim
\sum_{j=1}^N n_j$) with the restriction that the total angular momentum 
is $L$, i.e.,
\begin{equation}
\sum_{j=1}^N l_j + p N(N-1) = L ,
\end{equation}
where $n_j$ and $l_j$ are the $\Lambda$-level 
 and the angular momentum indices, respectively, for composite fermions.
In most of this paper, we choose states that have 
the lowest CF kinetic energy (zeroth order 
CF diagonalization).  By allowing hybridization with
CF states with higher kinetic energies (higher 
order CF diagonalization), more accurate approximations can be obtained, as 
seen explicitly below.  As illustrated in the previous section, 
the wave function $\Psi_\alpha^L$ corresponding to the $\alpha$th basis at
$L$ is then given by 
\begin{eqnarray} \nonumber
\Psi_\alpha^L &=& A \left[
\psi_{n_1^{(\alpha)},l_1^{(\alpha)}} (z_1) \cdot
\psi_{n_2^{(\alpha)},l_2^{(\alpha)}} (z_2) 
\cdots
\psi_{n_N^{(\alpha)},l_N^{(\alpha)}} (z_N)
\right] 
\\
&& \times \Phi_1^{2p}, \quad \quad \quad \quad (\alpha=1,2,\ldots, D^*)   
\end{eqnarray}
where
\begin{equation}
\psi_{n,l} (z_i) \equiv J_i^{-p} {\cal P}_{\rm LLL}[ \eta_n^l (z_i) J_i^p ].
\end{equation}

\begin{figure}
\epsfig{file=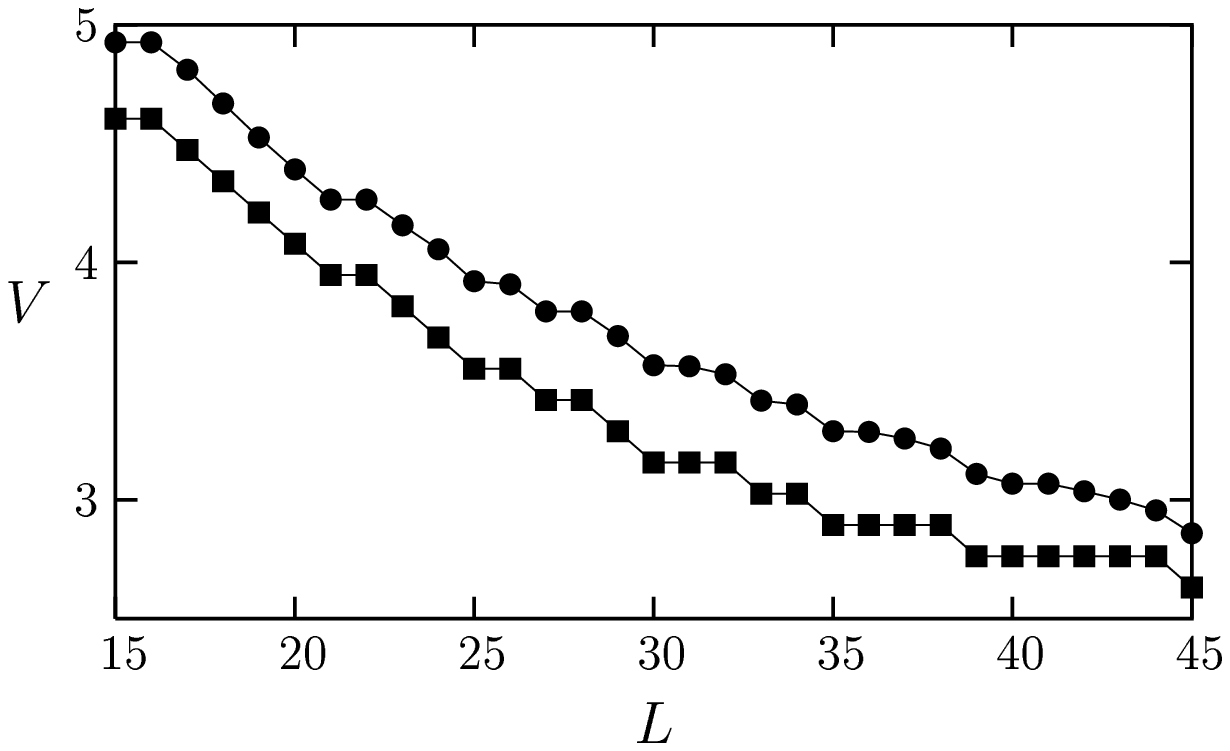,width=3.0in,angle=0}\\
(a)\\[5mm]
\epsfig{file=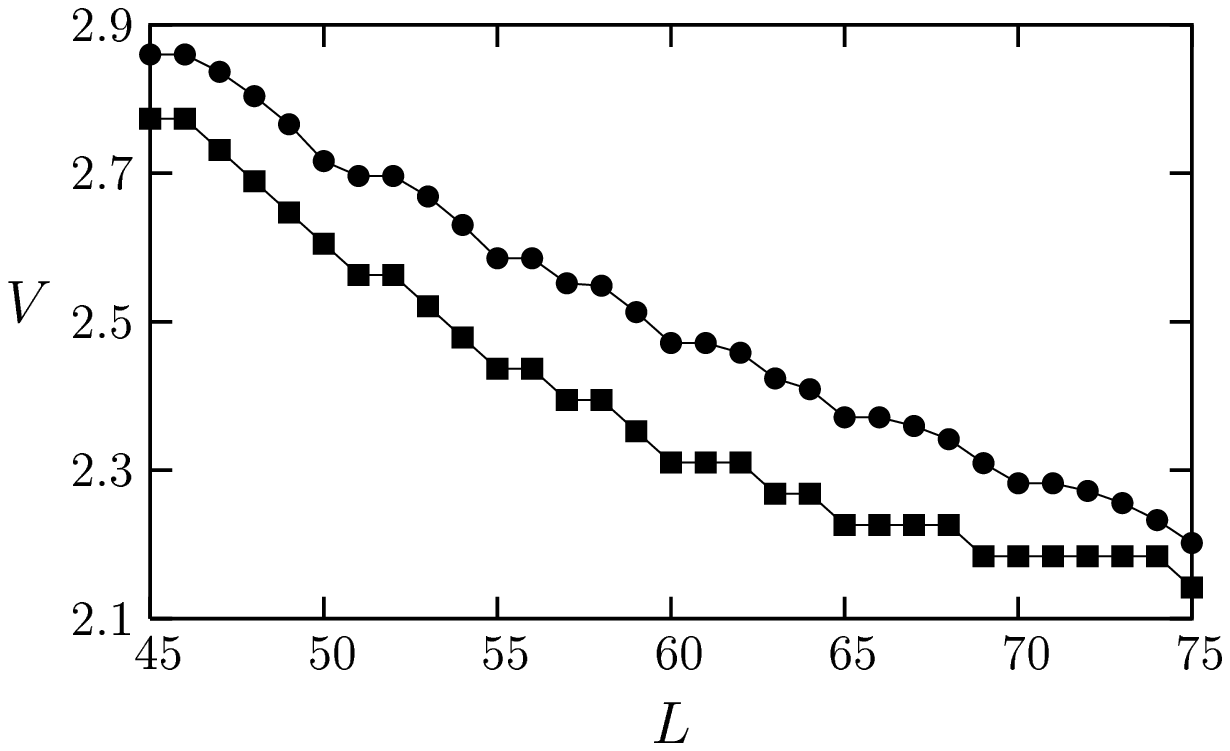,width=3.0in,angle=0}\\
(b)\\[5mm]
\epsfig{file=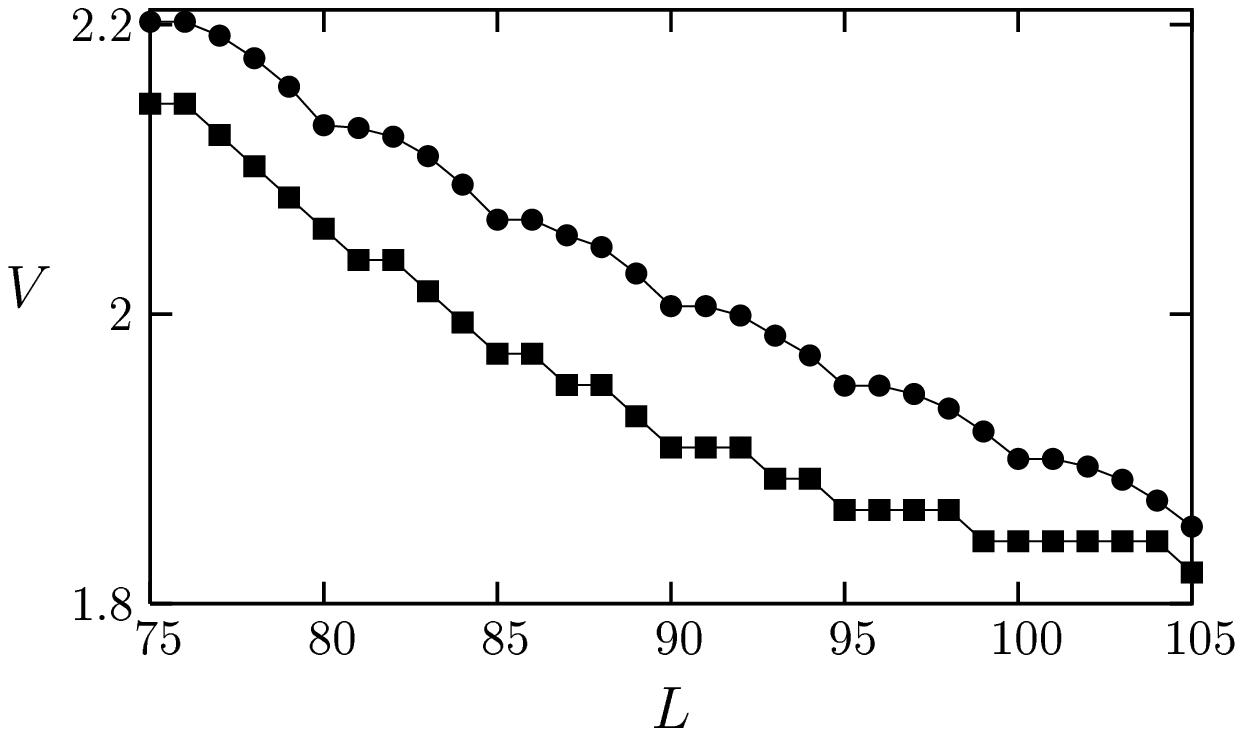,width=3.0in,angle=0}\\
(c)
\caption{\label{fig:mfcf}
	The exact interaction energy (circle) vs. the mean-field CF ground-state
energy (square) as a function of the angular momentum $L$ for $N=6$.
The exact energy is plotted in units of $e^2/\epsilon \ell$ 
while the CF energy is in arbitrary units.
}
\end{figure}
\begin{figure}
\epsfig{file=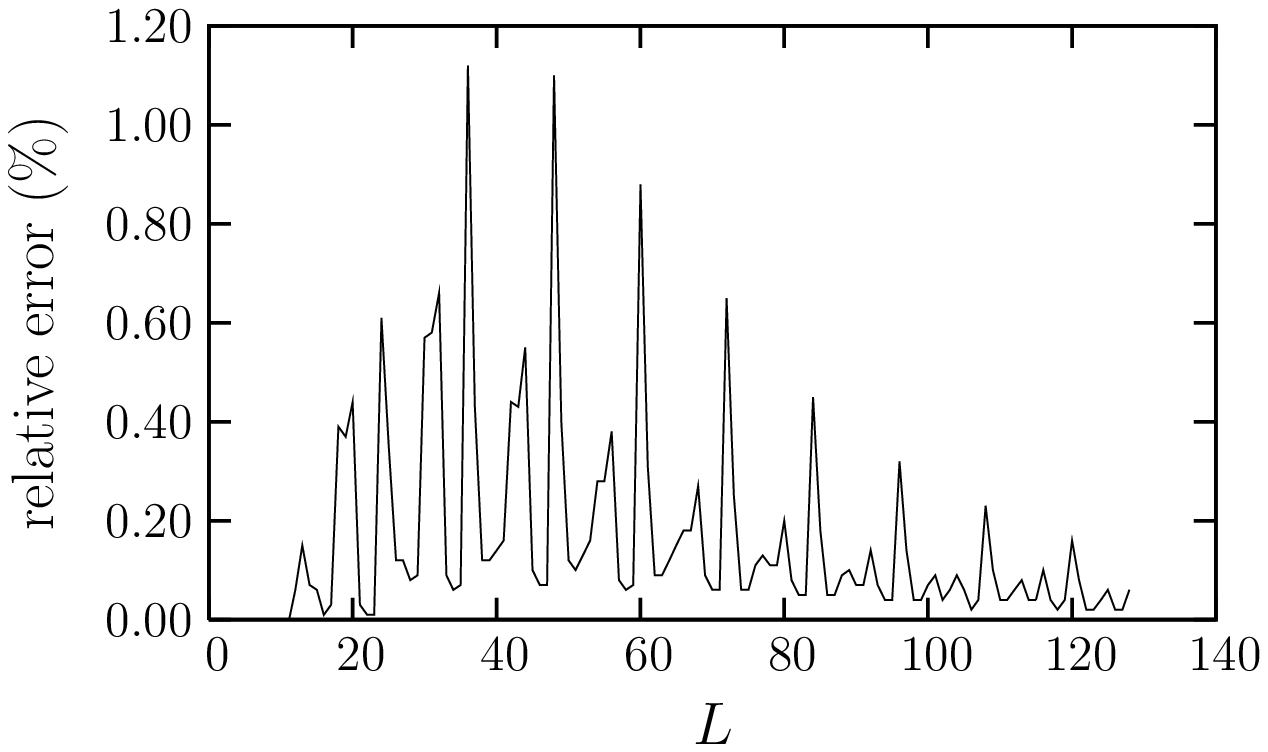,width=3.0in,angle=0}\\
(a)\\[2mm]
\epsfig{file=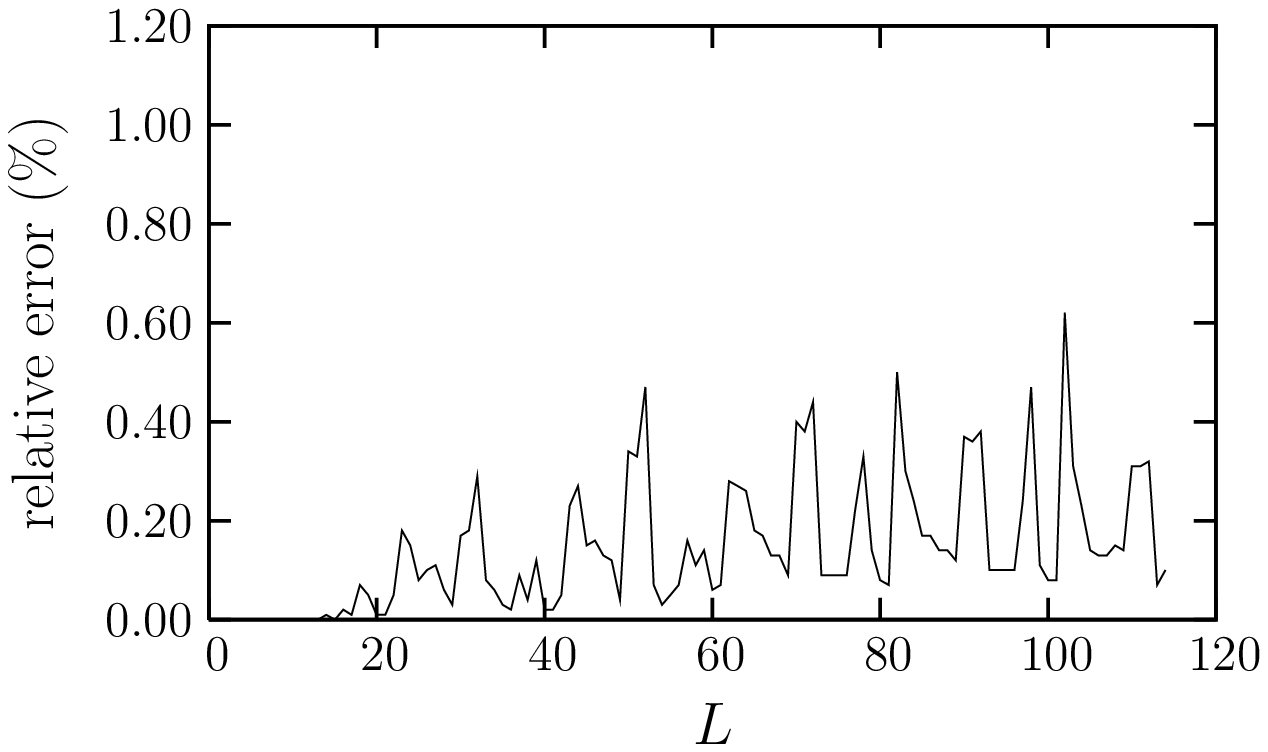,width=3.0in,angle=0}\\
(b)\\[2mm]
\epsfig{file=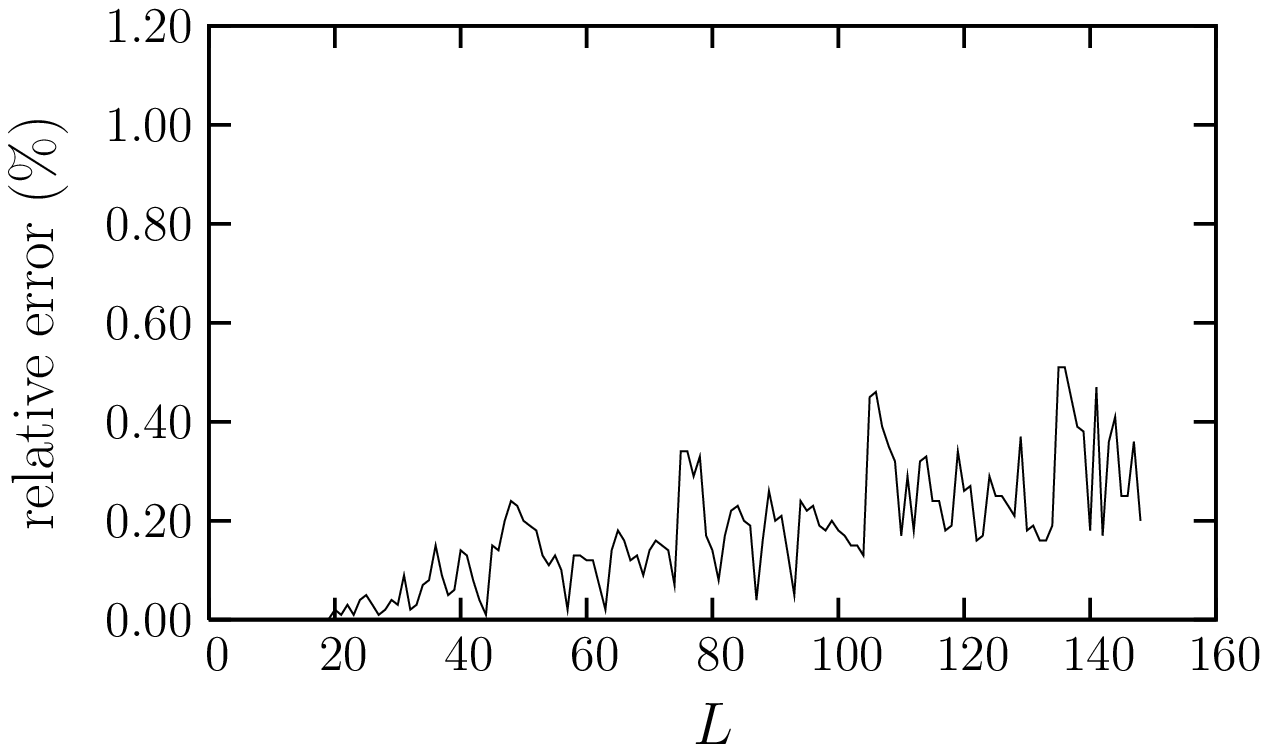,width=3.0in,angle=0}\\
(c)\\[2mm]
\epsfig{file=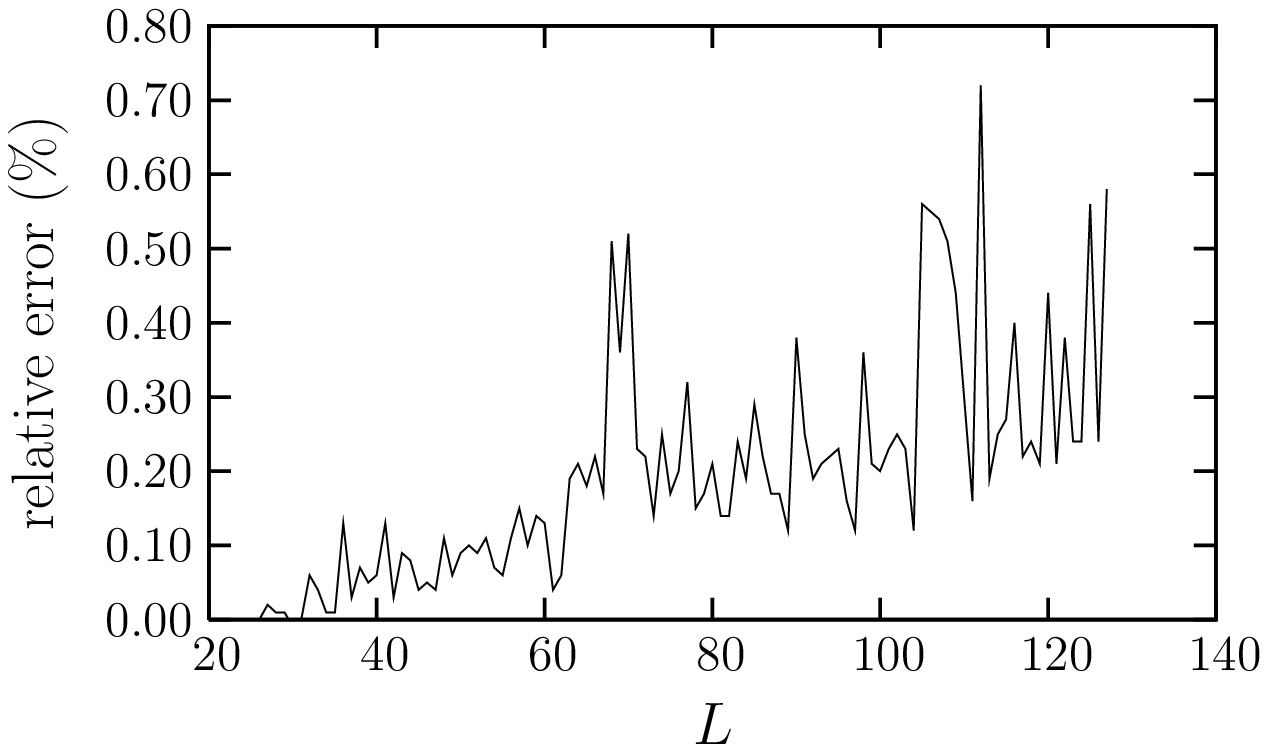,width=3.0in,angle=0}\\
	(d)
\caption{\label{fig:error}
	The relative errors of CF ground-state energy compared with the exact
energies for (a) $N=4$; (b) $N=5$; (c) $N=6$; (d) $N=7$.
}
\end{figure}

\begin{figure*}
\parbox{0.45\textwidth}{
\centerline{\epsfig{file=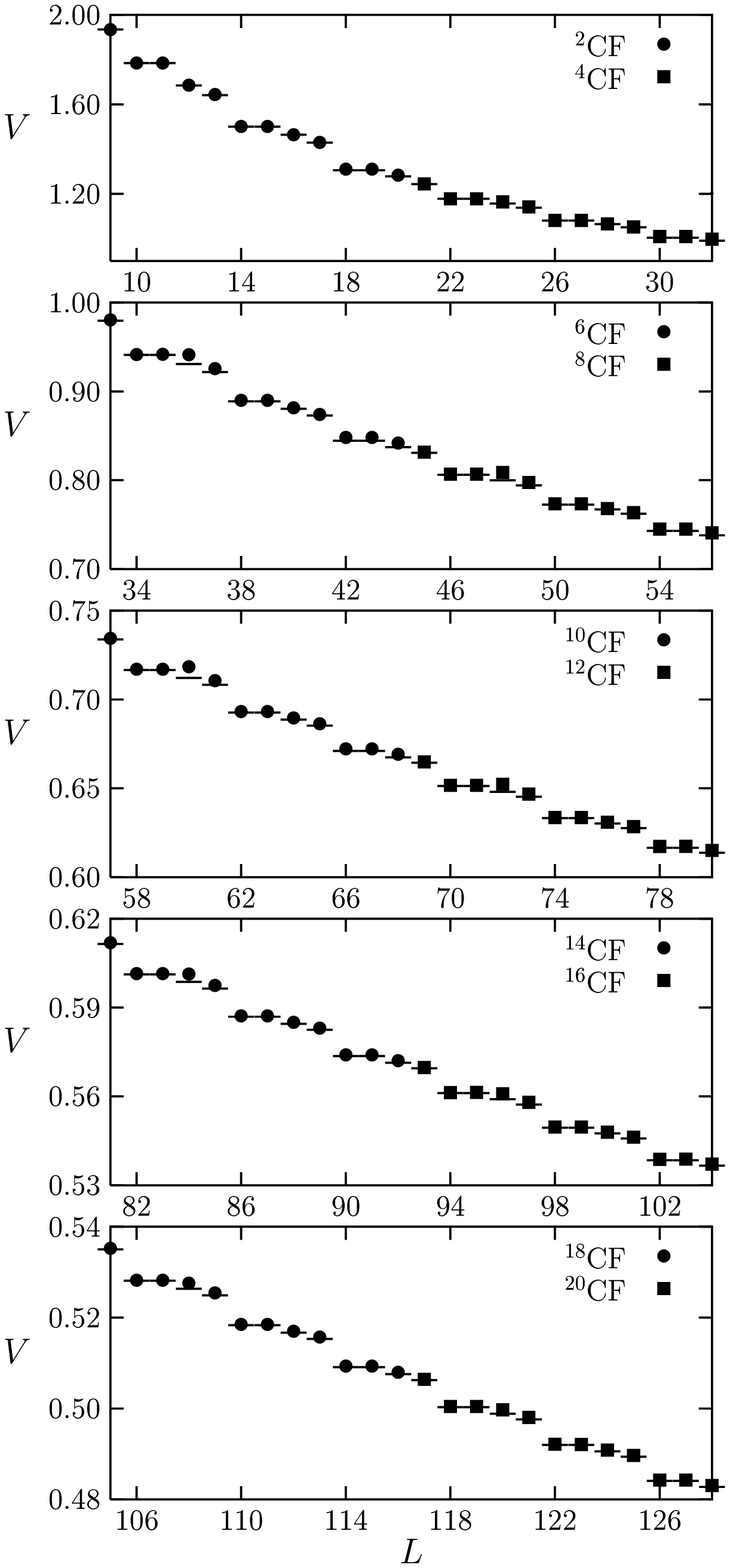,width=3.0in,angle=0}}
\centerline{(a)}
}
\hspace*{0.05\textwidth}
\parbox{0.45\textwidth}{
\centerline{\epsfig{file=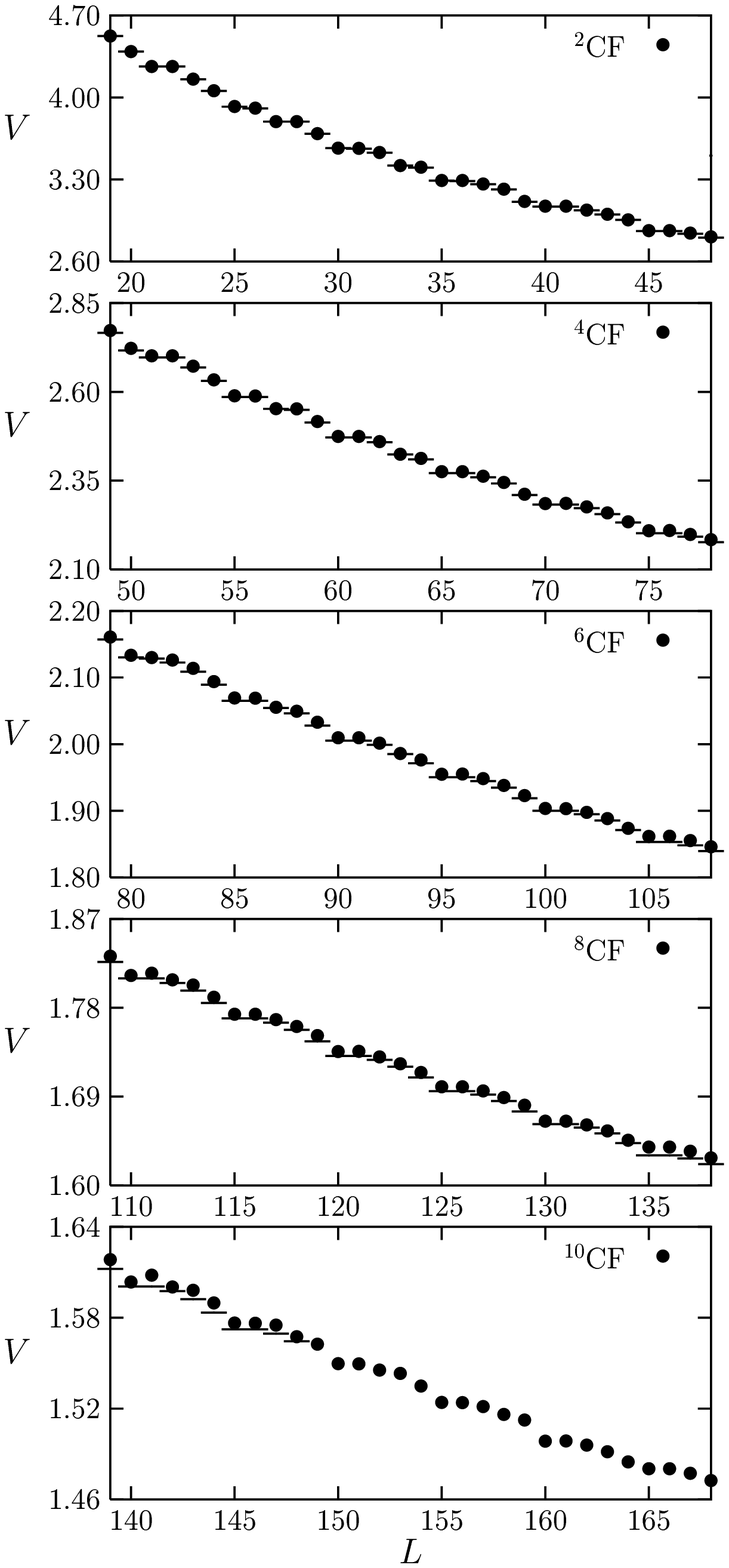,width=3.0in,angle=0}}
\centerline{(b)}
}
\caption{
The exact interaction energies $V$ (dashes) are given as a function of the angular momentum $L$ for (a) $N=4$ (b) $N=6$ particles.  
The dots are the predictions of the zeroth-order CF theory.
Different panels correspond to $L$ regions where composite fermions of 
different flavors are relevant. 
The energies are quoted in units of $e^2/\epsilon \ell$.  
}
\label{fig:Energy}
\end{figure*}

The CF basis states $\Psi_\alpha^L$ constructed in this manner 
are not orthogonal to one
another at given $L$, and sometimes not even linearly independent. 
We use Gram-Schmidt procedure to orthogonalize the states.
The orthogonal basis states $|\xi \rangle$ are expressed as
\begin{equation} \label{GSrel}
|\xi_\alpha \rangle = |\eta_\alpha \rangle
- \sum_{\gamma=1}^{\alpha-1} \frac{\langle \xi_\gamma | \eta_\alpha \rangle}
{\langle \xi_\gamma | \xi_\gamma \rangle}
| \xi_\gamma \rangle,
\end{equation}
where the normalized state $| \eta_\alpha \rangle$ is defined by
\begin{equation}
| \eta_\alpha \rangle \equiv 
\frac{| \Psi_\alpha \rangle}{\sqrt{\langle \Psi_\alpha |\Psi_\alpha \rangle}}.
\end{equation}
From the relation in Eq.~(\ref{GSrel}) we can find the recursion
relation for the transformation matrix $U_{\alpha\beta}$, defined by 
$|\xi_\alpha \rangle \equiv \sum_\beta U_{\alpha\beta} |\eta_\beta \rangle$,
\begin{equation}
U_{\alpha\beta} = \left\{
\begin{array}{cc}
\displaystyle - \sum_{\gamma=1}^{\alpha-1} 
\frac{ \sum_{\delta=1}^\gamma U_{\gamma\delta}^* {\cal
O}_{\delta\alpha} }
{\sum_{\delta,\epsilon=1}^\gamma U_{\gamma\delta}^* U_{\gamma\epsilon} {\cal O}_{\delta\epsilon}}
U_{\gamma\beta} 
& \hbox{for }  \beta < \alpha , \\
1 & \hbox{for } \beta = \alpha , \\
0 & \hbox{for } \beta > \alpha ,
\end{array}
\right.
\end{equation}
where ${\cal O}_{\alpha\beta} \equiv \langle \eta_\alpha | \eta_\beta
\rangle$. 
The computation of $U_{\alpha\beta}$ enables us to calculate the
Coulomb Hamiltonian matrix elements $V_{\alpha\beta}$ 
in orthonormal basis sets,
\begin{equation}  \label{Velement}
V_{\alpha\beta} =
\frac{
\langle \xi_\alpha | V | \xi_\beta \rangle
}{
\sqrt{
\langle \xi_\alpha | \xi_\alpha  \rangle
\langle \xi_\beta  | \xi_\beta \rangle
}
},
\end{equation}
where
\begin{eqnarray}
\langle \xi_\alpha | V | \xi_\beta \rangle
&=& \sum_{\gamma,\delta} U_{\alpha\gamma}^* U_{\beta\delta} 
\langle \eta_\gamma | V | \eta_\delta \rangle ,
\\
\langle \xi_\alpha | \xi_\alpha  \rangle
&=& \sum_{\gamma,\delta} U_{\alpha\gamma}^* U_{\beta\delta} 
{\cal O}_{\gamma\delta} .
\end{eqnarray}
The Coulomb Hamiltonian matrix in Eq.~(\ref{Velement}) is diagonalized 
to obtain the energy and the wave function of the ground state in the
CF theory.

The matrix elements ${\cal O}_{\alpha\beta}$ and 
$\langle \eta_\alpha | V | \eta_\beta \rangle$ have been evaluated
by the Metropolis Monte Carlo method.
We have performed in access of $4\times10^6$ Monte Carlo steps for 
each energy, which correctly gives five
significant digits for $N\le7$ and four significant digits for $N\ge8$.
The error bar denotes the standard deviation obtained 
from four independent runs.

\section{Mean-field model}
It is customary to first consider the mean-field version of CF model,
in which the interaction between composite fermions is assumed to vanish. 
Then the interaction energy of electron systems is completely transformed into
the ``kinetic energy'' of composite fermions. and
can be evaluated in units of CF cyclotron energy $\hbar \omega^{*}$ 
by summing the $\Lambda$-level indices of all occupied CF states
\begin{equation}
V_{\rm CF-MF} = \left(\sum_{j=1}^N n_j \right) \hbar \omega^* .
\end{equation}

Figure~\ref{fig:mfcf} compares the interaction energies 
predicted by the mean-field CF
theory with the exact ground-state energies as a function of total angular momentum
$L$ for $N=6$.  For small angular momenta 
the mean-field CF model predicts correctly the positions of cusps on
ground-state energy ({\it e.g.} see Fig.~\ref{fig:mfcf}(a)).
However, with increasing $L$, it gradually fails to capture the correct 
positions of cusp states.
Such discrepancies between the exact results and the mean-field predictions 
motivate the need for an investigation of the residual interactions between composite
fermions, considered in the subsequent sections.

\begin{figure}
\centerline{\epsfig{file=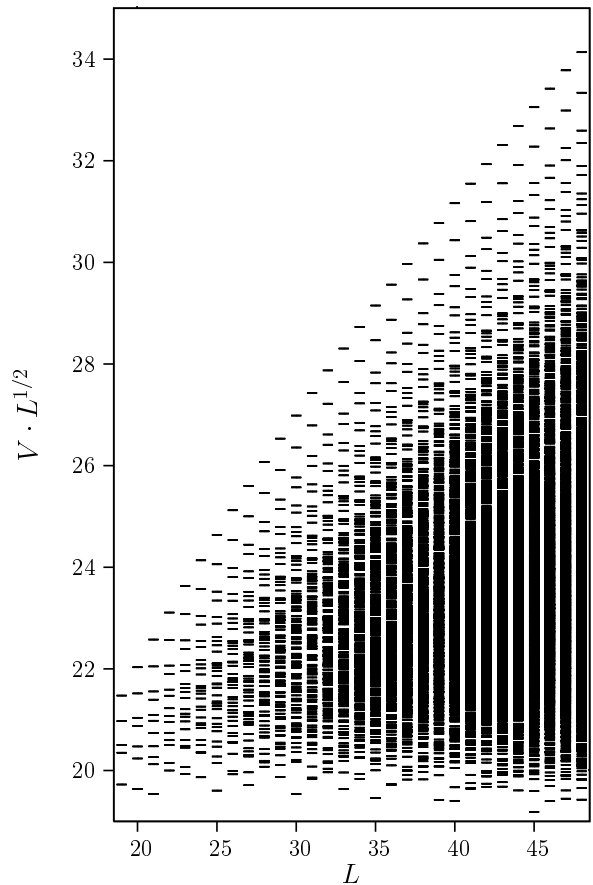,width=0.45\textwidth,angle=0}}
\caption{ \label{fig:fullex}
	Exact energy spectrum for $N=6$.
}
\end{figure}

\begin{figure*}
\epsfig{file=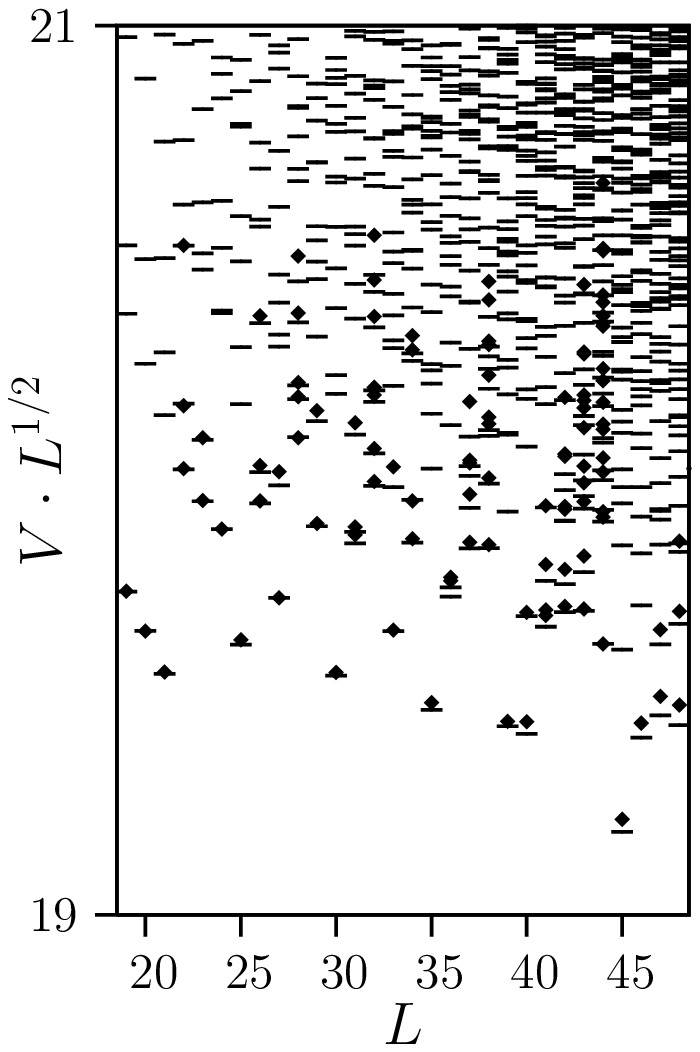, width=0.3\textwidth}
\epsfig{file=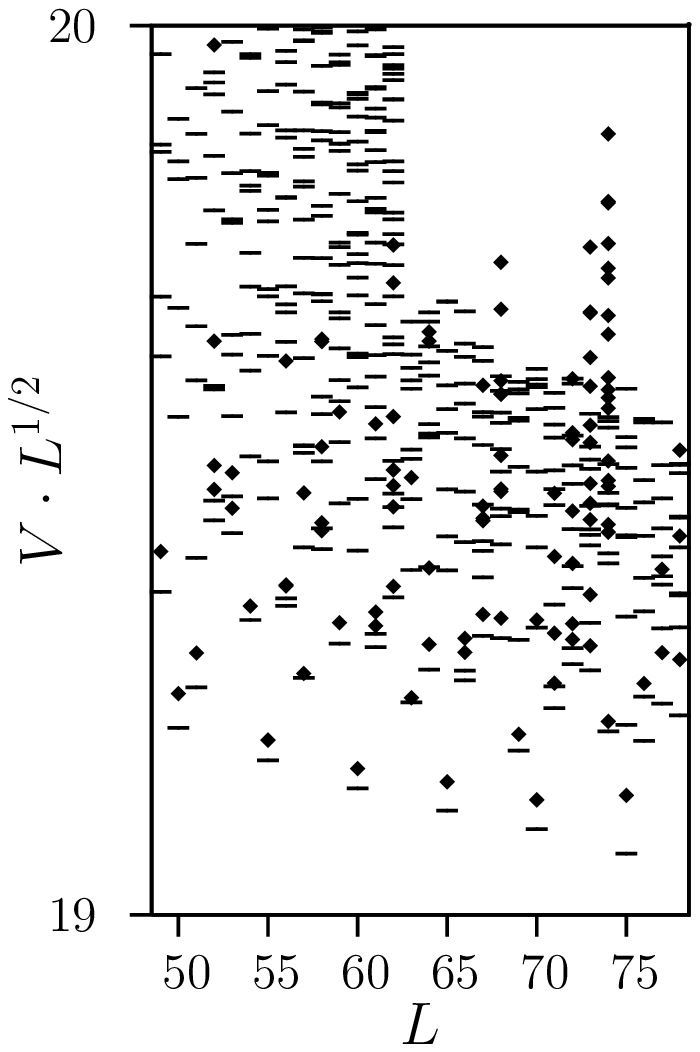, width=0.3\textwidth}
\epsfig{file=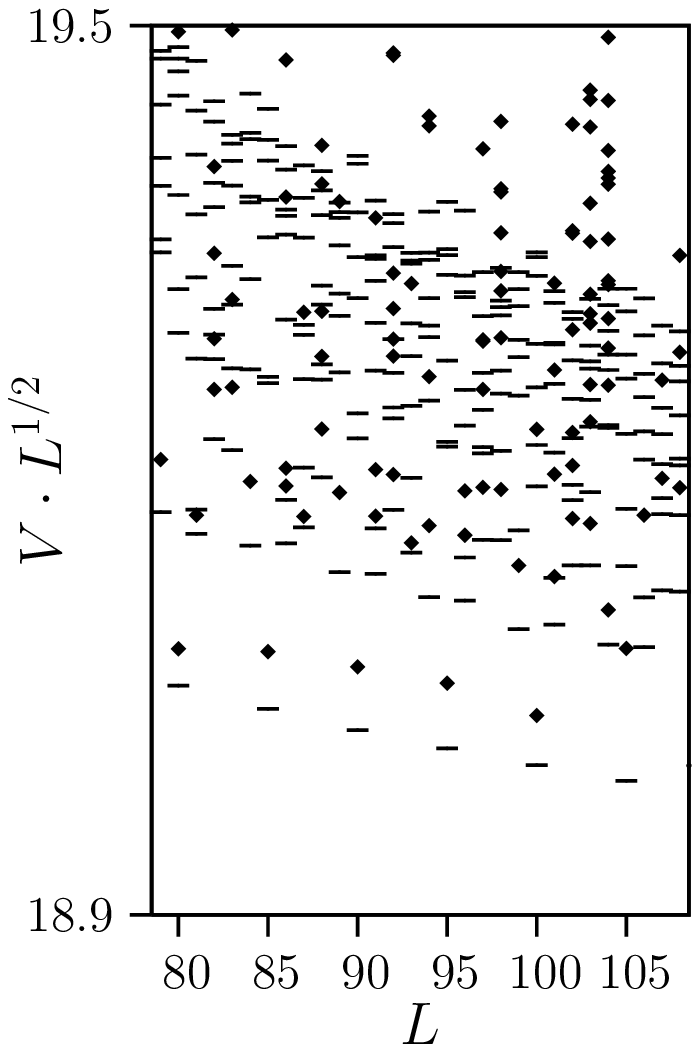, width=0.3\textwidth}
\caption{\label{fig:ex-CF0}
Exact energies (dashes) and the CF energies (dots) for six particles 
as a function of the angular momentum $L$.  The left, center, and right panels 
correspond to $^2$CFs, $^4$CFs, and $^6$CFs, respectively.  The CF 
energies are obtained in the zeroth order calculation (explained in text).  
For $L>62$, the exact spectrum is truncated because the 
energies are obtained by the Lanczos method.  The energies are multiplied by 
$\sqrt{L}$ for convenience of illustration.
}
\end{figure*}

\begin{figure*}
\epsfig{file=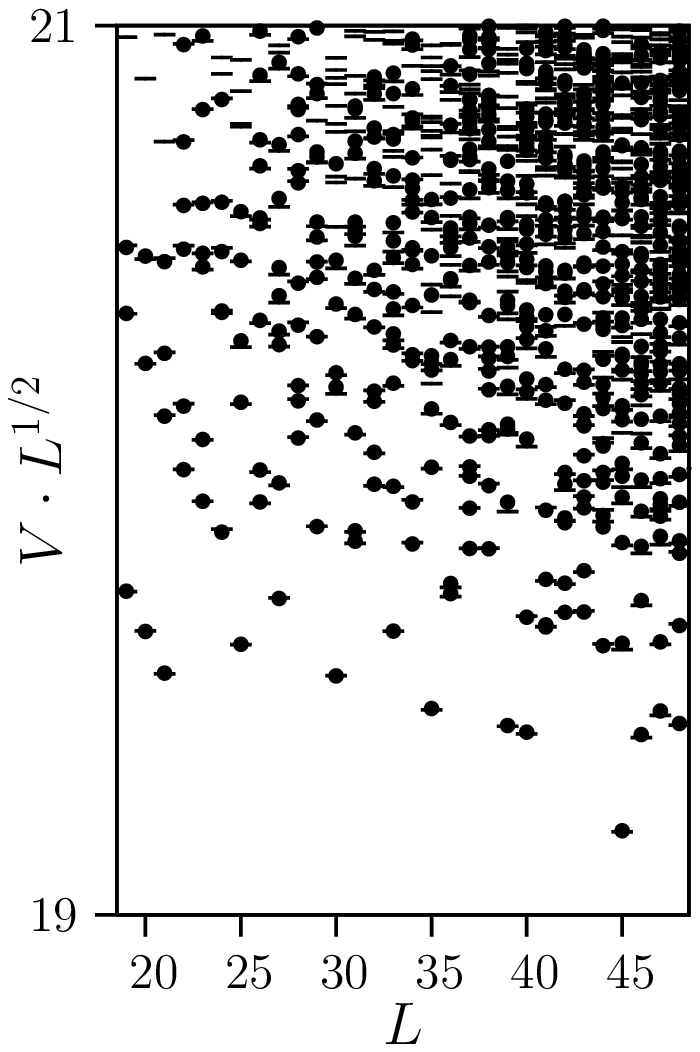, width=0.3\textwidth}
\epsfig{file=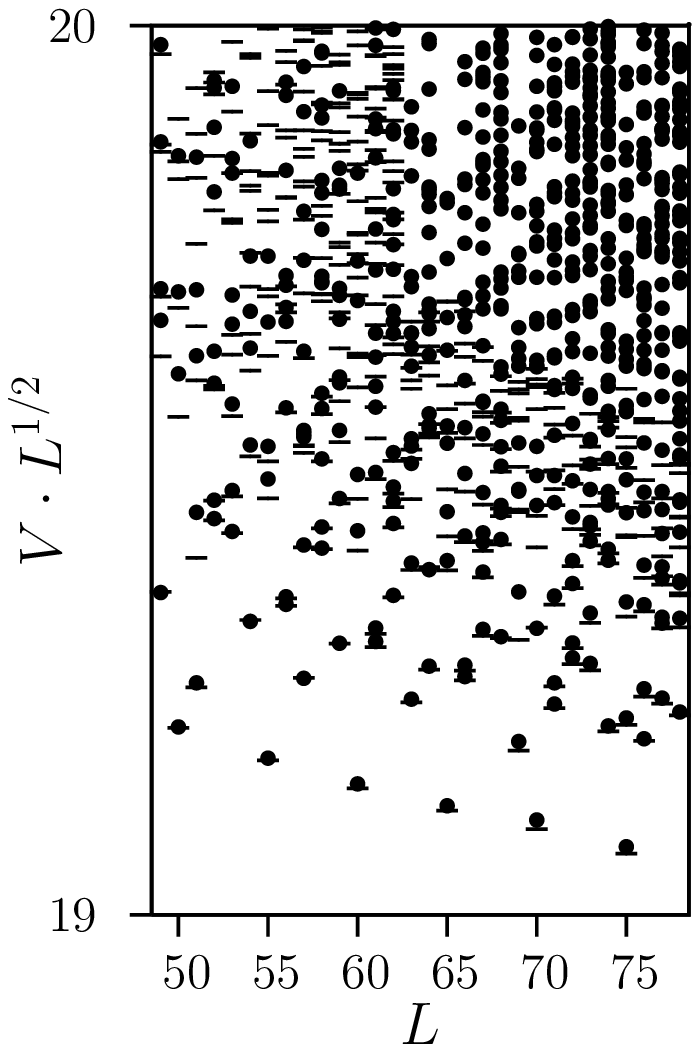, width=0.3\textwidth}
\epsfig{file=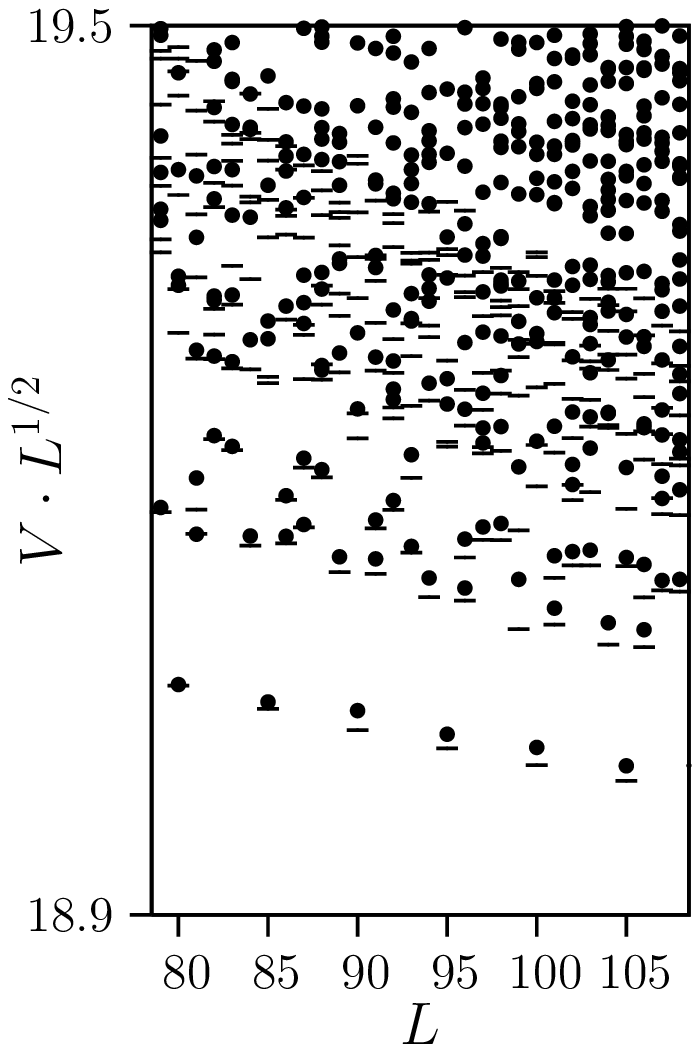, width=0.3\textwidth}
\caption{\label{fig:ex-CF1}
Same as in Fig. \ref{fig:ex-CF0}, but with the CF energies obtained in the 
first order calculation.
}
\end{figure*}

\section{Zeroth-order composite fermion theory}

Based on the CF diagonalization procedure described in
Sec.~\ref{subsec:CFdiag}
we have computed the CF ground state energies at the zeroth order, i.e. 
by including only the lowest kinetic energy CF basis states.
We have also carried out exact diagonalization for $N\leq 10$. 
It is possible to go to arbitrarily large values of $L$
within the CF theory, but available computer memory and execution time 
restricts our {\em exact} diagonalization study to $L \leq 148, 127, 118, 116$, 
and 113 for $N=6$, 7, 8, 9, and 10, respectively. 
(To our knowledge,~\cite{onefifth,Tsiper2,hole} 
the largest systems for which exact results have been
 computed are $N=8,L=140$ with $D_{\rm ex}=3023010$ and 
 $N=10,L=135$ with $D_{\rm ex}=2977866$.)
The results are listed in Tables \ref{tab:E4}--\ref{tab:E10},
which show that the CF energies compare well to the exact ones.
(Many of the exact energies have been reported previously in the 
literature,~\cite{Kamilla,Landman,CFcrystal,onefifth} but are included here for completeness.) 
The deviation between the CF and the exact energies ($V_{\rm CF}$ and $V_{\rm ex}$) is $L$ dependent but small in the entire range of $L$ studied in this work. 
The largest deviations are 1.1\% ($L=36,N=4$), 0.6\% ($L=102,N=5$), 
0.5\% ($L=135,N=6$), 0.7\% ($L=112,N=7$), 0.9\% ($L=89,N=8$), 
1.3\% ($L=114,N=9$), and 0.2\% ($L=80,N=10$).  For $N=4$, the maximum 
deviation occurs for composite fermions carrying 6-8 vortices;
for $N\ge5$, the deviation grows generally with $L$ in the
range considered in this work (see Fig.~\ref{fig:error}).

The plot of interaction energy as a function of $L$ in
Fig.~\ref{fig:Energy} gives a demonstration of the accuracy of the CF predictions.
In addition to the quantitative accuracy, 
the qualitative features of the energy versus $L$ plot 
are reproduced faithfully by the CF theory.
For $N=4$, the major cusps occurs at $L=4n+2$.
The system with six particles exhibits more complicated features.
For system of $^2$CFs (top panel) shows cusps at
$L=21,25,27,30,33,35,39,40$ and $45$.
As the flavor increases, the cusps at angular momenta 
other than $L=5n$ become less prominent and eventually disappear. 
Such periodic behavior is consistent with the geometric
interpretation.~\cite{Ruan,Landman}

\section{Next order CF theory}

The CF theory allows a systematic perturbative way of improving the results.
Above we considered only the CF states with the lowest CF kinetic energy, called 
the zeroth-order CF theory. The next step is to include states with one more 
unit of the kinetic energy in the CF basis.~\cite{Jeon}  The degree of 
improvement can be seen in Figs.~\ref{fig:fullex}, \ref{fig:ex-CF0}, and \ref{fig:ex-CF1}.
Figure~\ref{fig:fullex} shows the full spectrum for six electrons in a range of angular momentum; 
Figure~\ref{fig:ex-CF0} shows the spectrum from the zeroth order CF
diagonalization, and Fig.~\ref{fig:ex-CF1}
from the first order.  Both the zeroth and first order spectra capture the qualitative 
behavior and the positions of the cusps, but the first order theory is 
quantitatively much more accurate.  In both cases, the discrepancy between the 
CF and the exact energies ($V_{\rm CF}$ and $V_{\rm ex}$) grows with $L$, because while the dimension of the 
CF basis remains the same when $L$ is changed by $N(N-1)$, the dimension 
of the full lowest LL Fock space increases rapidly.  In spite of the small basis 
(which sometimes contains only one state at the zeroth order), the CF theory 
is quantitatively satisfactory.

\section{Concluding remarks}

Several authors~\cite{Seki,Landman} have 
noted that the CF theory fails to produce the cusp positions 
for large $L$.  These comparisons, however, 
refer only to the mean-field model of composite fermions,
in which the interaction energy of electrons at $L$ is
viewed as the kinetic energy of {\em free} fermions at an
effective angular momentum $L^*$, with the cyclotron energy
treated as a parameter.  The validity of the CF theory 
should not be confused with the validity of the mean-field picture, 
which serves, at best, as a useful guide; given its crudeness,
it is in fact surprising it works as well as it does.
A more substantive, microscopic test of the CF theory
requires working with the correlated wave functions produced by
the CF theory. 

Our extensive study of quantum dot states shows that 
the microscopic composite fermion theory, defined through 
wave functions, gives an excellent description in regions 
including both liquid-like and crystal-like ground states, and 
continues to be satisfactory from very low angular momenta 
to the largest angular momenta studied to date.   
It provides an accurate  approximation for the ground state wave function 
and the ground state energy at every single $L$ in the 
wide range studied, and correctly reproduces all cusps 
in plot of the ground state energy vs. $L$.  Taken together, these 
results constitute a detailed verification for  
the validity of the composite fermion theory for quantum dots.

It is expected, from general considerations,
that the ground state at large $L$ will 
resemble a classical crystal, because large $L$ implies small density
(or small filling factor) with particles far from one another, 
as a result of which the system behaves more or less classically. 
Reference~\onlinecite{Landman} has studied a Hartree-Fock crystal trial
wave function based on an analogy to the classical crystal
ground state in a quantum dot.  No crystalline correlations are put in by hand 
in our calculations described above, however.  As implied by the 
successful comparisons with the exact results, and also confirmed by 
an explicit calculation of the pair correlation function \cite{Jeon,Jeon2}, 
the CF theory automatically generates a crystal of correct symmetry.
The CF approach offers many other advantages over the Hartree-Fock electron crystal 
description.  The latter obtains wave functions and
energies only for certain special values of $L$, and even then only 
for the ground state.  The CF theory, on the other hand, provides a 
quantitative understanding of states at all $L$. 
For $N=6$, Reference~\onlinecite{Landman} explicitly quotes 
energies from their approach for seven values of $L$ in the 
range $75\leq L \leq 135$.  For these angular momenta, 
the zeroth order CF theory gives lower energy in every case except 
at $L=135$.  (The state at $L=135$ has one filled 
$\Lambda$ level, described by the Laughlin wave function.)  At the first order,
the CF results improve substantially for large $L$.  As shown in 
Ref.~\onlinecite{CFcrystal},
an almost exact description of the crystallite at large $L$ is obtained in terms 
of a crystal of composite fermions, wherein a combination of the crystal 
and CF physics are introduced right at the outset.
It is gratifying that the principle that applies to the fractional quantum Hall effect in 
a bulk two-dimensional electron system also produces an understanding 
of the quantum dot physics at high magnetic fields.

Partial support by the National Science Foundation under grant
no. DMR-0240458 is gratefully acknowledged.


\begin{thebibliography}{99}
\bibitem{review}
For reviews on quantum dots and their possible applications, see
L.P. Kouwenhoven, G. Sch\"{o}n, and L.L. Sohn, in {\em Mesoscopic
Transport}
NATO ASI Series E (Kluwer Academic, Boston, 1997), Vol. 345;
G. Burkard and D. Loss, in {\em Semiconductor Spintronics and Quantum
	Computation},
	edited by D.D. Awschalom, D. Loss, and N. Samarth (Springer-Verlag,
			New York, 2002), pp. 230-276;
S.M. Reimann and M. Manninen, Rev. Mod. Phys.  {\bf 74}, 1283 (2002).

\bibitem{Yoshioka} D. Yoshioka, B.I. Halperin, and P.A. Lee, 
Phys. Rev. Lett. {\bf 50}, 1219 (1983).

\bibitem{Girvin} S.M. Girvin and T. Jach, Phys. Rev. B {\bf 28}, 4506 
(1983).

\bibitem{Dev} G. Dev and J.K. Jain, Phys. Rev. B {\bf 45}, 1223
(1992).

\bibitem{Beenakker93} C.W.J. Beenakker and B. Rejaei,
Physica B {\bf 189}, 147 (1993).

\bibitem{Yang} S.-R. Eric Yang, A.H. MacDonald, and M.D. Johnson, Phys.
Rev. Lett. {\bf 71}, 3194 (1993).

\bibitem{Tapash} P.A. Maksym and T. Chakraborty, Phys. Rev. Lett. {\bf
65}, 108 (1990).

\bibitem{Xie} X.C. Xie, S. Das Sarma, and S. He, Phys. Rev. B
 {\bf 47}, 15942 (1993).

\bibitem{Hawrylak} P. Hawrylak, Phys. Rev. Lett. {\bf 71}, 3347 (1993).

\bibitem{Kawamura} J.K. Jain and T. Kawamura, Europhys. Lett. {\bf
	29}, 321 (1995).

\bibitem{Kamilla} J.K. Jain and R.K. Kamilla, Int. J. Mod. Phys. B {\bf
	11}, 2621 (1997). 

\bibitem{Seki} T. Seki, Y. Kuramoto, and T. Nishino, J. Phys. Soc.
Jpn. {\bf 65}, 3945 (1996).

\bibitem{Manninen} M. Manninen, S. Viefers, M. Koskinen, and S.M. Reimann, 
Phys. Rev. B {\bf 64}, 245322 (2001). 

\bibitem{Cappelli} 
A. Cappelli, C. Mendez, J. Simonin, and G.R. Zemba, Phys. Rev. B {\bf 58},
16291 (1998); J.H. Han  and S.-R. Eric Yang, Phys. Rev. 
B {\bf 58}, R10163 (1998).

\bibitem{Landman} C. Yannouleas and U. Landman, Phys. Rev. B {\bf 68}, 
035326 (2003).

\bibitem{Ruan} W.Y. Ruan, Y.Y. Liu, C.G. Bao, and Z.Q. Zhang, Phys.
Rev. B {\bf 51}, R7942 (1995);
W.Y. Ruan and H.-F. Cheung, J. Phys.: Condens. Matter {\bf 11}, 435
(1999).

\bibitem{Maksym} P.A. Maksym, Phys. Rev. B {\bf 53}, 10871 (1996).

\bibitem{Harju} A. Harju, S. Silj\"amaki, and R.M. Nieminen,
Phys. Rev. Lett. {\bf 88}, 226804 (2002); 

\bibitem{RQLC} S.M. Reimann, M. Koskinen, Y. Yu, and M.  Manninen, 
	New J. Phys. {\bf 8}, 59 (2006).

\bibitem{Rei2} M. Koskinen, S.M. Reimann, J.-P. Nikkarila, and M.  
Manninen, cond-mat/0605321.

\bibitem{Toreblad} M. Toreblad, M. Borgh, M. Koskinen, M. Manninen, and
S.M. Reimann, Phys. Rev. Lett. {\bf 93}, 090407 (2004).

\bibitem{Cnell}  A.D.\ G\"{u}\c{c}l\"{u} J.-S. Wang and H. Guo, 
Phys. Rev. B {\bf 68}, 245323 (2002); 
A.D. G\"{u}\c{c}l\"{u} and C.J.\ Umrigar, {\it ibid.} {\bf 72}, 045309 (2003); 
A.D.\ G\"{u}\c{c}l\"{u}, G.S. Jeon, C.J. Umrigar, and J.K. Jain, 
{\it ibid.} {\bf 72}, 205327(2005); 
G.S. Jeon, A.D. G\"{u}\c{c}l\"{u}, C.J. Umrigar, and J.K. Jain, 
{\it ibid.} {\bf 72}, 245312 (2005).

\bibitem{Rei5} S.M. Reimann, M. Koskinen, M. Manninen, and B.R.
Mottelson, Eur. Phys. J. D  {\bf 16}, 381 (2001).

\bibitem{Peeters2}
M.B. Tavernier, E. Anisimovas, and F.M. Peeters,
Phys. Rev. B {\bf 74}, 125305 (2006).


\bibitem{Muller} H.-M. M\"uller and S.E. Koonin, Phys. Rev. B {\bf
54}, 14532 (1996).

\bibitem{Expt1} 
B. Su, V.J. Goldman, and J.E. Cunningham, Science {\bf
255}, 313 (1992); Phys. Rev. B {\bf 46}, 9644 (1992).

\bibitem{Expt2} R.C. Ashoori, H.L. Stormer, J.S. Weiner,
L.N. Pfeiffer, K.W. Baldwin, and K.W. West, Phys. Rev. Lett. {\bf 71},
613 (1993); Phys. Rev. Lett. {\bf 68}, 3088 (1992); 
R. C. Ashoori, Nature (London) {\bf 379}, 413 (1996). 

\bibitem{Expt3} B. Meurer, D. Heitman, and K. Ploog, Phys. Rev.
Lett. {\bf 68}, 1371 (1992).

\bibitem{Expt4} F. Findeis, M. Baier, A. Zrenner, M. Bichler, G.
Abstreiter, U. Hohenester, and E. Molinari, 
Phys. Rev. B {\bf 63}, 121309 (2001).

\bibitem{Tsui} D.C. Tsui, H.L. Stormer, and A.C. Gossard, Phys. Rev.
Lett. {\bf 48}, 1559 (1982).

\bibitem{Jeon} G.S. Jeon, C.-C. Chang, and J.K. Jain,
	Phys. Rev. B {\bf 69}, 241304(R) (2004).

\bibitem{Jeon2} G.S. Jeon, C.-C. Chang, and J.K. Jain,
	J. Phys.: Condens. Matter {\bf 16}, L271 (2004).

\bibitem{CFcrystal}
C.-C. Chang, G.S. Jeon, and J.K. Jain,
	Phys. Rev. Lett. {\bf 94}, 016809 (2005).

\bibitem{onefifth}
C.-C. Chang, C. T\"oke, G.S. Jeon, and J.K. Jain,
	Phys. Rev. B {\bf 73}, 155323 (2006).


\bibitem{Jain} J.K. Jain, Phys. Rev. Lett. {\bf 63}, 199 (1989);
Physics Today {\bf 53} (4), 39 (2000); {\em Composite Fermions,} in press 
(Cambridge University Press).

\bibitem{Mandal} S.S. Mandal and J.K. Jain, 
Phys. Rev. B {\bf 66}, 155302 (2002).

\bibitem{QP}
M. Kasner and W. Apel, Phys. Rev. B {\bf 48}, 11435 (1993):
U. Girlich and M. Hellmund, {\it ibid.} {\bf 49}, R17488 (1994):
V. Melik-Alaverdian and N.E. Bonesteel, {\it ibid.} {\bf 58}, 1451 (1998);
G.S. Jeon and J.K. Jain, {\it ibid.} {\bf 68}, 165346 (2003).


\bibitem{Girvin2}
S.M. Girvin and T. Jach, Phys. Rev. B {\bf 29}, 5617 (1984).

\bibitem{Stone} M. Stone, H. W. Wyld, and R. L. Schult, Phys. Rev. B {\bf 45}, 14156 (1992).

\bibitem{Tsiper} E.V. Tsiper, J. Math. Phys. {\bf 43}, 1664 (2002).

\bibitem{Gagliano} E.R. Gagliano, E. Gagotto, A. Moreo, and F.C. Alcaraz,
  Phys. Rev. B {\bf 34}, 1677 (1986).


\bibitem{Tsiper2} E.V. Tsiper and V.J. Goldman, Phys. Rev. B {\bf 64}, 165311
(2001).

\bibitem{hole}
G.S. Jeon and J.K. Jain, Phys. Rev. B {\bf 71}, 045337 (2005).


\end{thebibliography}
\end{document}